\documentclass[]{aastex631}

\usepackage{newtxtext}
\usepackage{newtxmath}
\usepackage[T1]{fontenc}
\usepackage{graphicx}	
\usepackage{amsmath}	
\usepackage{amssymb}	
\usepackage{amsfonts}
\usepackage{xcolor}
\usepackage{multirow}
\usepackage{subfigure}
\usepackage{hyperref}
\usepackage{float}
\usepackage{booktabs}    
\usepackage{tabularx}    
\usepackage{threeparttable} 
\usepackage{ragged2e}
\makeatletter
\let\newfloat\newfloat@ltx
\makeatother
\usepackage{algorithm}
\usepackage{algpseudocode}

\begin{document}

\title{FRTSearch: Unified Detection and Parameter Inference of Fast Radio Transients using Instance Segmentation}

\correspondingauthor{ Xiaoyao Xie, Yabiao Wang, Pei Wang, and Di Li}
\email{ xyx@gznu.edu.cn, caseywang@tencent.com, wangpei@nao.cas.cn,  dili@mail.tsinghua.edu.cn,}

\author[0009-0009-2860-7836]{Bin Zhang}
\affiliation{School of Mathematical Sciences, Guizhou Normal University, Guiyang 550001, China;}
\affiliation{Guizhou Provincial Key Laboratory of Information and Computing Science , Guizhou Normal University, Guiyang 550001, China;}
\affiliation{School of Cyber Science and Technology, Guizhou Normal University, Guiyang 550001, China;}

\author[0000-0002-6592-8411]{Yabiao Wang}
\affiliation{Tencent Youtu Lab, Shanghai 201103, China}

\author{Xiaoyao Xie}
\affiliation{Guizhou Provincial Key Laboratory of Information and Computing Science , Guizhou Normal University, Guiyang 550001, China;}
\affiliation{School of Cyber Science and Technology, Guizhou Normal University, Guiyang 550001, China;}

\author{Shanping You}
\affiliation{Guizhou Provincial Key Laboratory of Information and Computing Science , Guizhou Normal University, Guiyang 550001, China;}
\affiliation{School of Cyber Science and Technology, Guizhou Normal University, Guiyang 550001, China;}

\author{Xuhong Yu}
\affiliation{Guizhou Provincial Key Laboratory of Information and Computing Science , Guizhou Normal University, Guiyang 550001, China;}
\affiliation{School of Cyber Science and Technology, Guizhou Normal University, Guiyang 550001, China;}

\author{Qiuhua Li}
\affiliation{Guizhou Provincial Key Laboratory of Information and Computing Science , Guizhou Normal University, Guiyang 550001, China;}
\affiliation{School of Cyber Science and Technology, Guizhou Normal University, Guiyang 550001, China;}

\author{Hongwei Li}
\affiliation{Guizhou Provincial Key Laboratory of Information and Computing Science , Guizhou Normal University, Guiyang 550001, China;}
\affiliation{School of Cyber Science and Technology, Guizhou Normal University, Guiyang 550001, China;}

\author{Shaowen Du}
\affiliation{Guizhou Provincial Key Laboratory of Information and Computing Science , Guizhou Normal University, Guiyang 550001, China;}
\affiliation{School of Cyber Science and Technology, Guizhou Normal University, Guiyang 550001, China;}

\author[0000-0002-9441-2190]{Chenchen Miao}
\affiliation{College of Physics and Electronic Engineering, Qilu Normal University, Jinan 250200, China}

\author[0000-0002-7420-9988]{Dengke Zhou}
\affiliation{Zhejiang Lab, Hangzhou, Zhejiang 311121, China}

\author[0000-0001-9956-6298]{Jianhua Fang}
\affiliation{Zhejiang Lab, Hangzhou, Zhejiang 311121, China}

\author[0000-0002-1036-5076]{Jiafu Wu}
\affiliation{Tencent Youtu Lab, Shanghai 201103, China}

\author[0000-0002-3386-7159]{Pei Wang}
\affiliation{National Astronomical Observatories, Chinese Academy of Sciences, Beijing 100101, China}
\affiliation{Institute for Frontiers in Astronomy and Astrophysics, Beijing Normal University, Beijing, 102206, China}
\affiliation{State Key Laboratory of Radio Astronomy and Technology, Chinese Academy of Sciences, A20 Datun Road, Chaoyang District, Beijing, 100101, China}

\author[0000-0003-3010-7661]{Di Li}
\affiliation{New Cornerstone Science Laboratory, Department of Astronomy, Tsinghua University, Beijing 100084, China}
\affiliation{National Astronomical Observatories, Chinese Academy of Sciences, Beijing 100101, China}
\affiliation{State Key Laboratory of Radio Astronomy and Technology, Chinese Academy of Sciences, A20 Datun Road, Chaoyang District, Beijing, 100101, China}

\begin{abstract}
The exponential growth of data from modern radio telescopes presents a significant challenge to traditional single-pulse search algorithms, which are computationally intensive and prone to high false-positive rates due to Radio Frequency Interference (RFI). In this work, we introduce FRTSearch, an end-to-end framework unifying the detection and physical characterization of Fast Radio Transients (FRTs). Leveraging the morphological universality of dispersive trajectories in time-frequency dynamic spectra, we reframe FRT detection as a pattern recognition problem governed by the cold plasma dispersion relation. To facilitate this, we constructed CRAFTS-FRT, a pixel-level annotated dataset derived from the Commensal Radio Astronomy FAST Survey (CRAFTS), comprising 2{,}392 instances across diverse source classes. This dataset enables the training of a Mask R-CNN model for precise trajectory segmentation.
Coupled with our physics-driven IMPIC algorithm, the framework maps the geometric coordinates of segmented trajectories to directly infer the Dispersion Measure (DM) and Time of Arrival (ToA).
Benchmarking on the FAST-FREX dataset shows that FRTSearch achieves a 98.0\% recall, competitive with exhaustive search methods, while reducing false positives by over 99.9\% compared to PRESTO and delivering a processing speedup of up to $13.9\times$. Furthermore, the framework demonstrates robust cross-facility generalization, detecting all 19 tested FRBs from the ASKAP survey without retraining. By shifting the paradigm from ``search-then-identify'' to ``detect-and-infer,'' FRTSearch provides a scalable, high-precision solution for real-time discovery in the era of petabyte-scale radio astronomy.
\end{abstract}


\section{Introduction} \label{sec:intro}

Pulsars \citep{Hewish_1968}, Rotating Radio Transients (RRATs) \citep{Mclaughlin_2006}, and Fast Radio Bursts (FRBs) \citep{Lorimer_2007} serve as natural laboratories for exploring fundamental physical laws under extreme conditions. Pulsars provide rigorous tests for general relativity \citep{hulse1975} and form the cornerstone of timing arrays for detecting nanohertz gravitational waves \citep{hobbs2010}, while polarization surveys further demonstrate their diagnostic power for probing magnetospheric geometry and the Galactic magnetic field \citep{han2006, zhanglei2025}. The intermittent radiation from RRATs offers a critical window into the magnetospheric activity of neutron stars \citep{Mclaughlin_2006}, while FRBs serve as novel probes for mapping cosmic baryon distribution and measuring intergalactic magnetic fields \citep{Zhang_2023}. Observationally, single-pulse emissions from these three classes of astrophysical sources share a defining characteristic: after propagating through the interstellar medium, they exhibit consistent dispersive trajectories on sub-second timescales, collectively termed Fast Radio Transients (FRTs). Consequently, systematically searching for and identifying FRTs, acting as observational beacons for these sources, is of paramount importance for advancing the frontiers of fundamental physics.

With the operation of modern radio telescopes such as the Five-hundred-meter Aperture Spherical radio Telescope (FAST) \citep{nan2011five}, the Canadian Hydrogen Intensity Mapping Experiment (CHIME), and the future Square Kilometre Array (SKA), the discovery rate of FRTs is growing exponentially. For instance, CHIME reported 536 FRBs in a single year \citep{amiri2021}, while FAST detected 1,652 bursts from FRB 20121102 in just 47 days \citep{li2021} and has discovered over 1,000 new pulsars to date \citep{lorimer2025}. The resulting petabyte-scale data deluge is pushing traditional search methods to their computational limits \citep{li_2018}, identifying data processing as the primary bottleneck constraining scientific output. Developing next-generation search algorithms capable of real-time, efficient FRT identification is therefore crucial to unlocking the full scientific potential of these new-era facilities.

\begin{figure}[htbp]
    \centering
    \includegraphics[width=\textwidth]{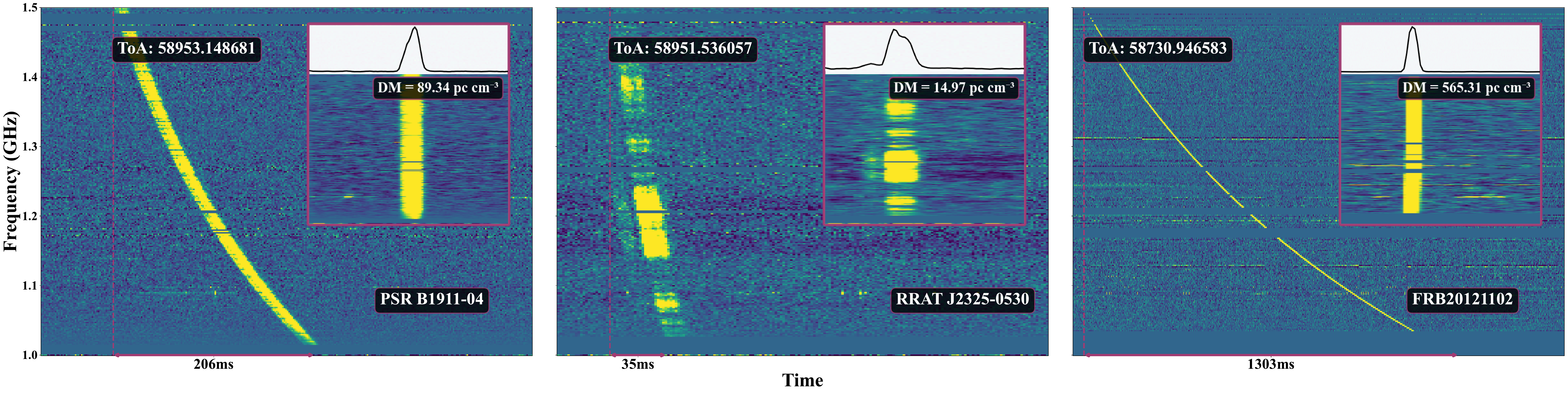}
    \caption{Waterfall plots for three typical source classes detected by the CRAFTS survey with the FAST telescope: PSR B1911$-$04 (left), RRAT J2325$-$0530 (middle), and FRB 20121102 (right). Each plot displays a time window beginning at the source's ToA. \textbf{Inset Panels:} The corresponding de-dispersed pulse profiles, obtained by aligning the signal in time using the source's DM.
    }
    \label{fig:waterfall_examples}
\end{figure}

When propagating through ionized media, astrophysical FRTs undergo frequency-dependent delays due to dispersion \citep{lorimer2005}. This effect transforms an intrinsically instantaneous pulse into a swept-frequency signal, manifesting as a characteristic parabolic curve in dynamic spectra (Figure~\ref{fig:waterfall_examples}). We term this signature the ``dispersive trajectory.'' Beyond serving as a critical diagnostic for rejecting Radio Frequency Interference (RFI), this trajectory encodes vital physical information quantified by the Dispersion Measure (DM)---the integrated column density of free electrons along the line of sight. Correcting for this dispersive delay via de-dispersion \citep{lorimer2005} is essential to recover the intrinsic temporal structure of the pulse, making accurate DM inference a prerequisite for scientific analysis.

Given the \emph{a priori} unknown observational parameters of FRTs, the prevailing detection paradigm universally employs a single-pulse search method via software packages such as PRESTO \citep{ransom2001}, Heimdall \citep{barsdell2012}, and TransientX \citep{men2024}. This approach necessitates a computationally intensive exhaustive search over a two-dimensional parameter space of trial DMs and pulse widths. The core process involves de-dispersing the time series across a range of trial DMs and subsequently convolving the results with matched filters to identify events with significant Signal-to-Noise Ratios (S/N) \citep{Cordes_2003}. Such brute-force methods suffer from high computational redundancy and generate numerous spurious candidates, predominantly triggered by RFI \citep{pang2018}. To mitigate this challenge, various automated candidate identification methods have been developed, evolving from early techniques scoring candidates against astrophysical priors \citep{karako2015, deneva2016} to classical machine learning classifiers \citep{devine2016, pang2018, michilli2018, pang2020}. Recently, deep learning techniques like Convolutional Neural Networks have demonstrated superior performance by classifying candidate diagnostic plots \citep{connor2018, agarwal2020, wang2024x}, while the latest methods employ object detection to directly localize FRBs in DM--time space \citep{zhang2025}. Despite these advances, candidate identification merely provides reactive mitigation for the massive volume of candidates, leaving the inherent inefficiency and redundancy of the search process itself unaddressed.

An alternative approach seeks to fundamentally circumvent the computational overhead of exhaustive searches by exploiting the characteristic dispersive trajectory of FRTs in time-frequency dynamic spectra (Figure~\ref{fig:waterfall_examples}) for direct pattern recognition. Early implementations targeted specific repeating FRB sources with known DMs, such as FRB 20121102 \citep{zhang2018} and FRB 20201124A \citep{liu2022}, where prior knowledge enabled tailored detection algorithms. Similar strategies have been applied to searches for Extraterrestrial Intelligence (ETI) and to identifying pulsars and FRBs \citep{gajjar2021, gajjar2022}. Recent work employs object detection networks to directly localize dispersive trajectories for FRB identification \citep{guo2024}. Although computationally efficient, these methods are predominantly designed for specific sources with prior information and lack the capability to directly infer critical parameters such as DM. This limitation restricts their utility as general-purpose discovery tools for large-scale blind surveys that must explore vast, unknown parameter spaces.

Instance segmentation, a computer vision technique that generates precise, pixel-level masks for individual object instances, provides a powerful analytical framework for extracting complex visual patterns from image data \citep{he2017mask}. In contrast to traditional object detection methods like R-CNN \citep{girshick2015fast} and YOLO \citep{redmon2016you}, which provide only coarse bounding box localizations, instance segmentation architectures exemplified by Mask R-CNN \citep{he2017mask} precisely delineate object boundaries. This approach has proven highly effective for the discovery and characterization of scientific targets across multi-modal astronomical data, including galaxy morphology classification \citep{burke2019, farias2020}, gravitational-wave detection \citep{aveiro2022, wang2024rapid}, and RFI mitigation \citep{akeret2017, gu2024radio}.

To address the fundamental limitations of existing paradigms, we propose a transformative approach motivated by a critical morphological observation (Figure~\ref{fig:waterfall_examples}): despite their diverse astrophysical origins, single pulses from pulsars, RRATs, and FRBs all exhibit morphologically consistent dispersive trajectories governed by the cold plasma dispersion relation \citep{lorimer2005}. This universality enables reframing FRT detection as a unified pattern recognition problem in time-frequency space. Our key insight is that pixel-level segmentation of dispersive trajectories simultaneously achieves two objectives: it enables precise signal localization without exhaustive parameter space searches, and it allows direct inference of physical parameters (particularly DM) by inverting the geometric information of segmented trajectories through the dispersion relation. This naturally motivates the application of instance segmentation techniques, which provide the pixel-level masks necessary to capture complete trajectory morphology for subsequent physics-driven parameter inference.

We present \texttt{FRTSearch\footnote{\url{https://github.com/BinZhang109/FRTSearch}}} (Fast Radio Transients Search), an end-to-end framework that synergizes deep learning-based instance segmentation with physics-driven parameter inference. By employing Mask R-CNN \citep{he2017mask} to directly localize and delineate dispersive trajectories in time-frequency dynamic spectra, followed by our Iterative Mask-based Parameter Inference and Calibration (IMPIC) algorithm, we derive DM and ToA directly from segmented trajectory coordinates. This unified architecture offers three distinct advantages: (i) it bypasses the computationally expensive DM-width grid searches required by traditional methods; (ii) it enables direct physical parameter inference, a capability absent in prior image-based approaches; and (iii) it seamlessly integrates detection and characterization within a single framework. This paradigm shift from ``search-then-identify'' to ``detect-and-infer'' represents a critical step toward the real-time, intelligent processing of massive data streams in modern radio astronomy. This paper is organized as follows: Section~\ref{sec:data} details the construction of our dataset. Section~\ref{sec:methodology} introduces the Mask R-CNN detection framework and the IMPIC algorithm. Section~\ref{sec:experiments} benchmarks performance against traditional methods and validates cross-facility generalization. Finally, Section~\ref{sec:limitations} discusses current limitations, and Section~\ref{sec:conclusions} concludes the paper.

\section{Dataset Construction}
\label{sec:data}

High-quality, pixel-level annotated data are a prerequisite for developing robust deep learning models in astronomical transient detection. In this section, we present the CRAFTS-FRT dataset, a pixel-level annotated dataset dedicated to FRTs. Comprising 2,392 instances (2,115 pulsars, 15 RRATs, and 262 FRBs), the dataset spans a broad range of observational parameters, with DM ranging from 14.9 to 780~pc~cm$^{-3}$ and S/N from 5.2 to 102.9. Leveraging the 19-beam receiver of the FAST telescope, the dataset captures natural morphological variations across diverse viewing geometries. This intrinsic diversity establishes CRAFTS-FRT as a standardized and challenging benchmark for instance segmentation tasks in radio astronomy.

\subsection{FAST Observational Data}
\label{sec:fast_data}

The data were acquired via the CRAFTS pulsar survey \citep{li_2018} using the FAST 19-beam L-band receiver, which covers a frequency range of 1,000--1,500~MHz with a total bandwidth of 500~MHz. Each observational file consists of 4,096 frequency channels uniformly sampled across the band (central frequency 1,250~MHz) and 131,072 time samples with a resolution of 98.304~$\mu$s, resulting in a total duration of approximately 12.88~s. Consequently, the raw time-frequency dynamic spectra possess dimensions of $4{,}096 \times 131{,}072$ pixels.

\begin{figure}[htbp]
    \centering
    \includegraphics[width=\textwidth]{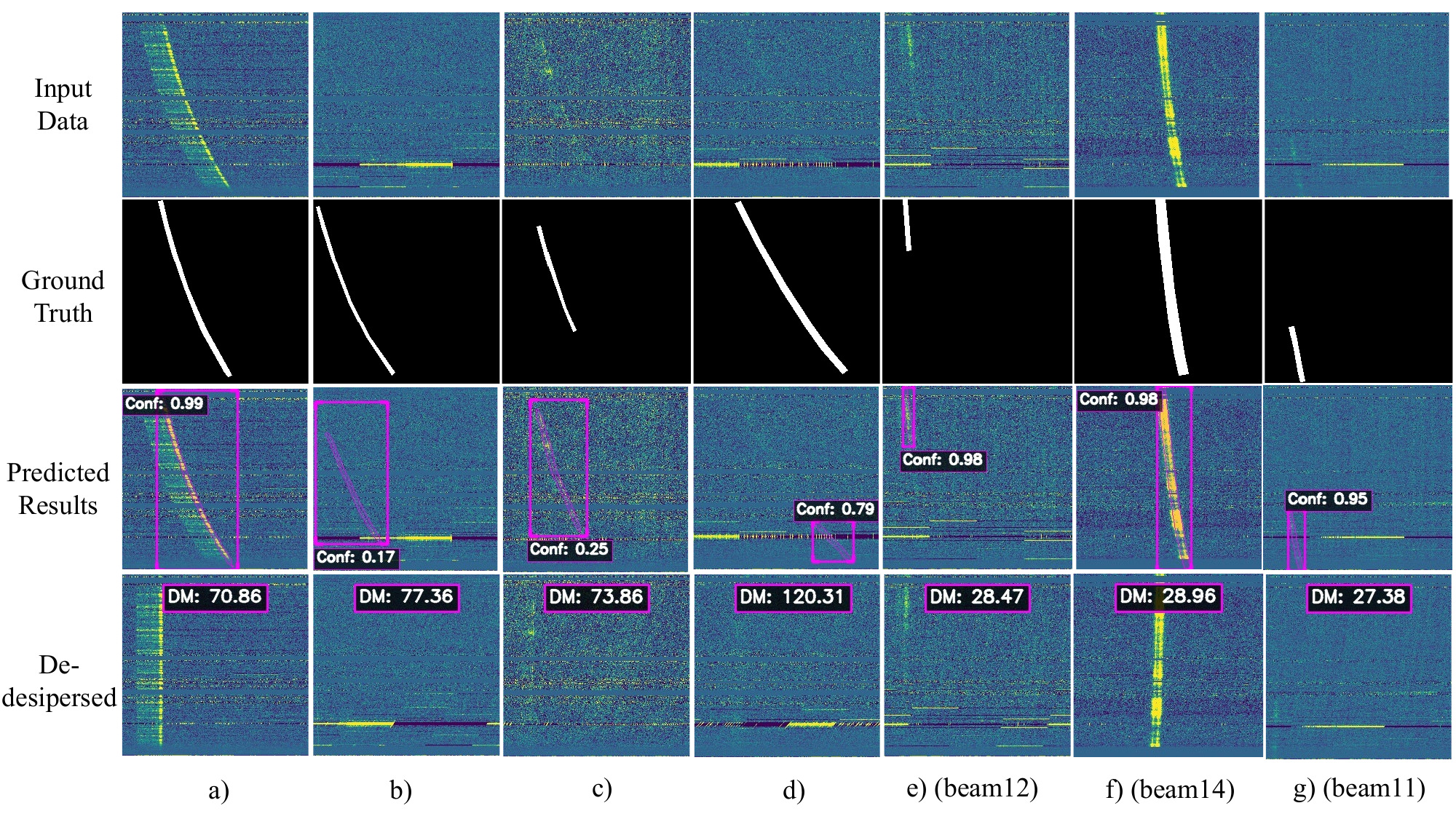}
    \caption{
        Visualization of key stages in the \texttt{FRTSearch} pipeline using seven representative examples (resized to $256 \times 256$). 
        \textbf{Columns (a)--(d):} Four distinct FRT sources exhibiting varied dispersive features.
        \textbf{Columns (e)--(g):} Pulsar B0820$+$02 observed simultaneously by three adjacent FAST receiver beams, demonstrating signal variation.
        \textbf{Row 1:} Preprocessed time-frequency dynamic spectra (model input).
        \textbf{Row 2:} Manually annotated ground-truth masks for training.
        \textbf{Row 3:} Mask R-CNN predictions displaying segmentation masks (pink contours), bounding boxes (pink rectangles), and confidence scores.
        \textbf{Row 4:} De-dispersed dynamic spectra with IMPIC-inferred DM values.
    }
    \label{fig:pipeline_overview}
\end{figure}

The multi-beam capability offers a critical advantage for dataset construction: it allows for the simultaneous detection of a single transient event across adjacent beams. This naturally generates samples with varying signal intensities and pulse morphologies due to beam sensitivity patterns (Figure~\ref{fig:pipeline_overview}, panels e--g), thereby enriching the dataset with intrinsic diversity essential for training robust and generalizable models.

\subsection{Data Preprocessing}
\label{sec:preprocessing}

The preprocessing pipeline converts raw observational data into standardized $256 \times 8,192$ matrices, optimized for model input (Figure~\ref{fig:pipeline_overview}, row 1). Adopting methodologies from the PRESTO package's \texttt{waterfaller.py}\footnote{\url{https://github.com/scottransom/presto/blob/master/bin/waterfaller.py}}, the pipeline proceeds through four distinct stages:

\begin{itemize}
    \item \textbf{Data Extraction:} Time-frequency dynamic spectra are extracted from the raw files, initializing the data matrices with dimensions of $4{,}096 \times 131{,}072$.
    
    \item \textbf{Downsampling:} A downsampling factor of 16 is applied along both frequency and time axes. This operation reduces the dimensions to a standardized $256 \times 8{,}192$, effectively enhancing the per-pixel S/N while reducing the computational load.
    
    \item \textbf{RFI Mitigation:} Narrowband and impulsive interference are identified and masked using statistical thresholding via PRESTO's \texttt{rfifind} routine \citep{ransom2002fourier}.
    
    \item \textbf{Normalization:} A two-step normalization strategy is implemented. First, per-channel Z-score normalization eliminates bandpass irregularities and standardizes the background noise floor. Subsequently, an exponential transformation is applied to maximize the contrast between the signal trajectories and the background noise.
\end{itemize}

This standardized protocol ensures consistent input formats across varying observational conditions while preserving the characteristic $\nu^{-2}$ dispersive trajectories geometry required for effective model training.

\subsection{Pixel-Level Annotation Strategy}

Training instance segmentation models requires precise pixel-level annotations to define trajectory boundaries accurately. We formulate this problem as a binary classification task, where each pixel in the $256 \times 8{,}192$ dynamic spectrum is labeled as either FRT signal (foreground) or background (noise and RFI). Following protocols established by large-scale computer vision benchmarks like COCO \citep{lin2014}, we manually annotated all 500 observation files, generating high-fidelity binary segmentation masks that serve as the Ground Truth (GT) for model supervision (Figure~\ref{fig:pipeline_overview}, row 2).

\subsection{Dataset Statistics and Splitting}

We curated 500 independent observation files from the CRAFTS survey to construct the dataset, which includes:

\begin{itemize}
    \item \textbf{Pulsars:} 291 files from 43 distinct pulsars, contributing 2,115 annotated instances (DM range: 21.0--780~pc~cm$^{-3}$).
    \item \textbf{RRATs:} 9 files from 3 RRATs, contributing 15 annotated instances (DM range: 14.9--29.8~pc~cm$^{-3}$).
    \item \textbf{FRBs:} 200 files (100 per source) from two repeating sources, FRB~20121102 (DM $\approx$ 565~pc~cm$^{-3}$;~\citealt{li2021}) and FRB~20201124A (DM $\approx$ 420~pc~cm$^{-3}$;~\citealt{zhang2022}), contributing 262 annotated instances.
\end{itemize}

\begin{figure}[htbp]
    \centering
    \includegraphics[width=1.0\textwidth]{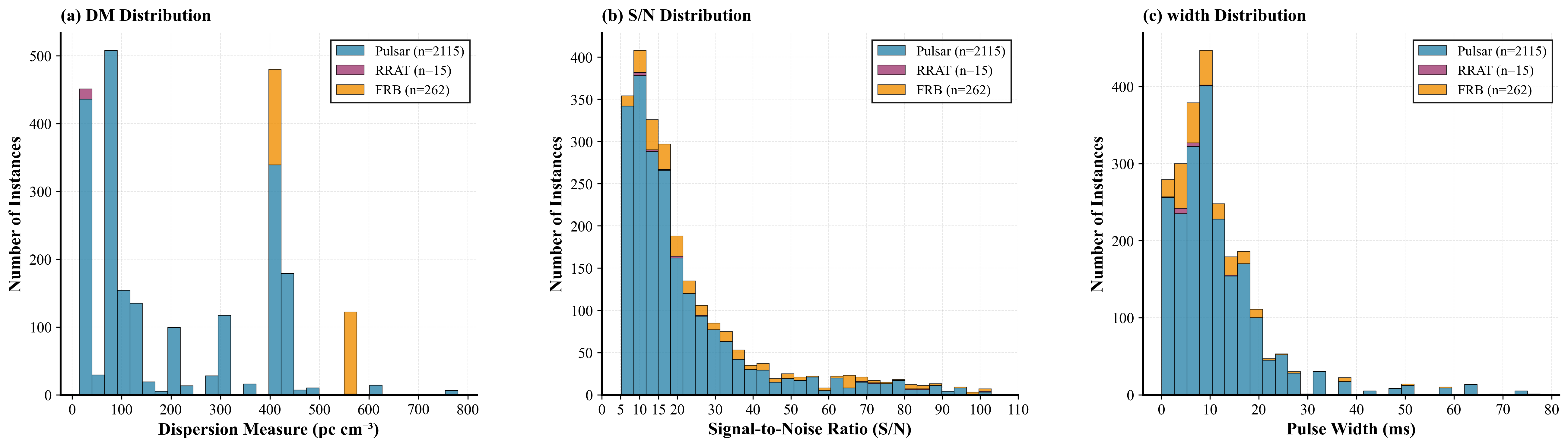}
    \caption{
        Distribution of observational parameters (DM, S/N, and pulse width) 
        across the full dataset of 2,392 FRT instances.
    }
    \label{fig:parameter_distributions}
\end{figure}

The complete dataset contains 2,392 pixel-level annotated FRT instances. To assess the parameter space coverage, we measured the S/N and pulse widths using TransientX \citep{men2024} (see Figure~\ref{fig:parameter_distributions}). The distribution covers S/N values from 5.2 to 102.9 and pulse widths from 0.1 to 77.7~ms. Notably, approximately 8.1\% of the instances exhibit S/N $< 7$, and 23.4\% exhibit S/N $< 10$. This inclusion of faint signals ensures the training set represents challenging observational conditions often missed by traditional search algorithms \citep{ransom2001, barsdell2012, men2024}.

For model development and evaluation, we employed an 80/20 train-validation split (Table~\ref{tab:dataset_summary}):

\begin{itemize}
    \item \textbf{Training Set:} 400 files (37 pulsars, 2 RRATs, and 80 files per FRB source), containing 1,805 instances.
    \item \textbf{Validation Set:} 100 files (6 pulsars, 1 RRATs, and 20 files per FRB source), containing 587 instances.
\end{itemize}

To prevent data leakage, Pulsar and RRAT sources were assigned exclusively to either the training or validation set, ensuring evaluation on unseen sources. For the two repeating FRBs, data splitting was performed at the file level ($80\%/20\%$). While the source DM remains constant, the stochastic nature of burst morphology, intensity, and background noise ensures that the validation set remains representative of unseen observational conditions. Detailed source information is provided in Table~\ref{tab:dataset_summary}.

\begin{table*}[!p]
\centering
\small
\caption{Detailed composition and train-validation splitting of the CRAFTS-FRT dataset.}
\label{tab:dataset_summary}
\begin{threeparttable}
    \setlength{\tabcolsep}{8pt}
    \begin{tabular*}{\textwidth}{@{\extracolsep{\fill}}rlcccccl@{}}
    \toprule
    \textbf{No.} & \textbf{Source Name} & \textbf{R.A. (hh:mm)} & \textbf{Decl. (dd:mm)} & \textbf{DM (pc\,cm$^{-3}$)} & \textbf{Files} & \textbf{FRT Instances} & \textbf{Set}\\
    \midrule
    1  & PSR B0559-05   & 06:01 & $-$05:27 &  80.5 &  17 & 278 & Training\\
    2  & PSR B0621-04   & 06:24 & $-$04:24 &  70.8 &  12 &  88 & Training\\
    3  & PSR B1841-05   & 18:44 & $-$05:38 & 411.7 &  10 & 299 & Training\\
    4  & PSR B1846-06   & 18:49 & $-$06:37 & 148.2 &   6 &  15 & Training\\
    5  & PSR J0458-0505 & 04:58 & $-$05:05 &  47.8 &  14 &  28 & Training\\
    6  & PSR J0459-0210 & 04:59 & $-$02:10 &  21.0 &  14 &  77 & Training\\
    7  & PSR J1814-0521 & 18:14 & $-$05:21 & 130.6 &  12 &  74 & Training\\
    8  & PSR J0640-00   & 06:39 & $-$00:04 &  64.8 &   1 &   1 & Training\\
    9  & PSR J1848-0511 & 18:48 & $-$05:11 & 418.0 &   8 &  40 & Training\\
    10 & PSR J0658+0022 & 06:58 & $+$00:55 & 115.6 &   1 &   3 & Training\\
    11 & PSR J1555-0515 & 15:55 & $-$05:15 &  23.5 &  10 &  51 & Training\\
    12 & PSR J1802-0523 & 18:02 & $-$05:23 & 121.0 &   3 &   4 & Training\\
    13 & PSR J1839-0459 & 18:39 & $-$04:59 & 237.0 &   7 &  13 & Training\\
    14 & PSR J1828-0611 & 18:28 & $-$06:11 & 363.2 &   1 &   6 & Training\\
    15 & PSR J1833-0559 & 18:33 & $-$05:59 & 346.7 &   2 &  10 & Training\\
    16 & PSR J1834-0602 & 18:34 & $-$06:02 & 445.0 &   9 &  83 & Training\\
    17 & PSR J1835-0522 & 18:35 & $-$05:22 & 456.0 &   1 &   7 & Training\\
    18 & PSR J1835-0600 & 18:35 & $-$06:00 & 780.0 &   4 &   6 & Training\\
    19 & PSR J1836-0517 & 18:36 & $-$05:17 & 564.0 &   1 &   1 & Training\\
    20 & PSR J1837-0559 & 18:37 & $-$05:59 & 319.5 &   3 &  17 & Training\\
    21 & PSR J1838-0453 & 18:38 & $-$04:53 & 617.2 &   2 &  14 & Training\\
    22 & PSR J1838-0549 & 18:38 & $-$05:49 & 276.6 &   5 &  19 & Training\\
    23 & PSR J1841-0524 & 18:41 & $-$05:24 & 284.5 &   1 &   9 & Training\\
    24 & PSR J1840-0559 & 18:40 & $-$05:59 & 319.1 &   9 &  35 & Training\\
    25 & PSR J1847-0605 & 18:47 & $-$06:05 & 207.9 &  12 &  57 & Training\\
    26 & PSR J1843-0459 & 18:43 & $-$04:59 & 444.1 &   9 &  96 & Training\\
    27 & PSR J1848-0545 & 18:45 & $-$05:45 & 315.9 &   1 &   1 & Training\\
    28 & PSR J1848-0601 & 18:48 & $-$06:01 & 496.6 &   2 &   6 & Training\\
    29 & PSR J1852-0635 & 18:52 & $-$06:36 & 173.9 &   3 &   5 & Training\\
    30 & PSR J1856-0525 & 18:56 & $-$05:26 & 131.8 &   8 &  57 & Training\\
    31 & PSR J1901-04   & 19:01 & $-$04:00 & 107.0 &   3 &   6 & Training\\
    32 & PSR J1910-0556 & 19:10 & $-$05:56 &  88.3 &   8 &  36 & Training\\
    33 & PSR J1921-05   & 19:21 & $-$05:23 &  80.7 &   8 &  29 & Training\\
    34 & PSR J1921-0510 & 19:21 & $-$05:10 &  96.6 &  10 &  44 & Training\\
    35 & PSR J2008+3758 & 20:07 & $+$37:58   & 143.0 &   2 &   4 & Training\\
    36 & PSR J2019+3810 & 20:19 & $+$38:09   & 495.0 &   2 &   4 & Training\\
    37 & PSR J2346-0609 & 23:46 & $-$06:09 &  22.5 &  12 &  53 & Training\\
    38 & PSR B0820+02   & 08:20 & $+$02:00   &  23.7 &  26 & 255 & Validation\\
    39 & PSR B1911-04   & 19:13 & $-$04:40 &  89.4 &   7 &  77 & Validation\\
    40 & PSR J1820-0509 & 18:20 & $-$05:09 & 102.4 &   6 &  79 & Validation\\
    41 & PSR J0652-0142 & 06:52 & $-$01:42 & 116.3 &   5 &  22 & Validation\\
    42 & PSR J1948+2333 & 19:48 & $+$23:33 & 198.2 &   4 &  42 & Validation\\
    43 & PSR J1845-0545 & 18:45 & $-$05:45 & 315.9 &  10 &  64 & Validation\\
    \midrule
    1 & RRAT J0447-04   & 04:47 & $-$04:35 &  29.8 &   5 &  11 & Training\\
    2 & RRAT J0139+3336   & 01:39 & $+$33:36 &  21.2 &   2 &  2 & Training\\
    3 & RRAT J2325-0530& 23:25 & $-$05:30 &  15.0 &   2 &   2 & Validation\\
    \midrule
    1 & FRB 20121102   & 05:32 & +33:05 & 565.0   & 100 & 121 & Training/Validation\tnote{a} \\
    2 & FRB 20201124A  & 05:08 & +26:03 & 420.0   & 100 & 141 & Training/Validation\tnote{a} \\
    \bottomrule
    \end{tabular*}
    \begin{tablenotes}
    \item[a] For each FRB, 80 files (FRB 20121102: 97 instances; FRB 20201124A: 119 instances) were used for training, and 20 files (FRB 20121102: 24 instances; FRB 20201124A: 22 instances) for validation. \item \textbf{Note:} The dataset is publicly available at \url{https://doi.org/10.57760/sciencedb.Fastro.00038}. 
    \end{tablenotes}
\end{threeparttable}
\end{table*}

\section{Methodology}
\label{sec:methodology}
We present the \texttt{FRTSearch} framework, which unifies dispersive trajectory detection with physics-driven parameter inference. Diverging from traditional ``search-then-identify'' methods that rely on exhaustive blind de-dispersion, our approach adopts a ``detect-and-infer'' paradigm, treating FRT discovery as a pattern recognition problem within time-frequency dynamic spectra. This methodology is intrinsically grounded in the physics of signal propagation through the Interstellar Medium , where the dispersive delay creates a characteristic morphological signature. Specifically, the time delay $\Delta t$ between two frequency channels $\nu_{\text{high}}$ and $\nu_{\text{low}}$ is rigorously governed by the cold plasma dispersion relation \citep{lorimer2005}:
\begin{equation}
\Delta t \approx k_{\mathrm{DM}} \times \mathrm{DM} \times \Delta(\nu^{-2}), 
\quad \text{where} \quad \Delta(\nu^{-2}) = \nu_{\mathrm{low}}^{-2} - \nu_{\mathrm{high}}^{-2}
\label{eq:dispersion}
\end{equation}
Here, $k_{\mathrm{DM}} \approx 4.1488 \times 10^3\, \mathrm{MHz}^{-2}\, \mathrm{pc}^{-1}\, \mathrm{cm}^3\, \mathrm{s}$ represents the dispersion constant. Equation~\eqref{eq:dispersion} establishes a deterministic mapping between the physical parameters and the trajectory geometry: given the pixel coordinates $\{(t_i, \nu_i)\}$ of a signal, the DM and ToA are uniquely encoded in the curvature and vertex of the trajectory. Consequently, the \texttt{FRTSearch} framework is designed to first extract these coordinates via instance segmentation and subsequently invert Equation~\eqref{eq:dispersion} to infer the physical parameters.

\subsection{Framework Overview}

\begin{figure}[htbp]
    \centering
    \includegraphics[width=1.0\textwidth]{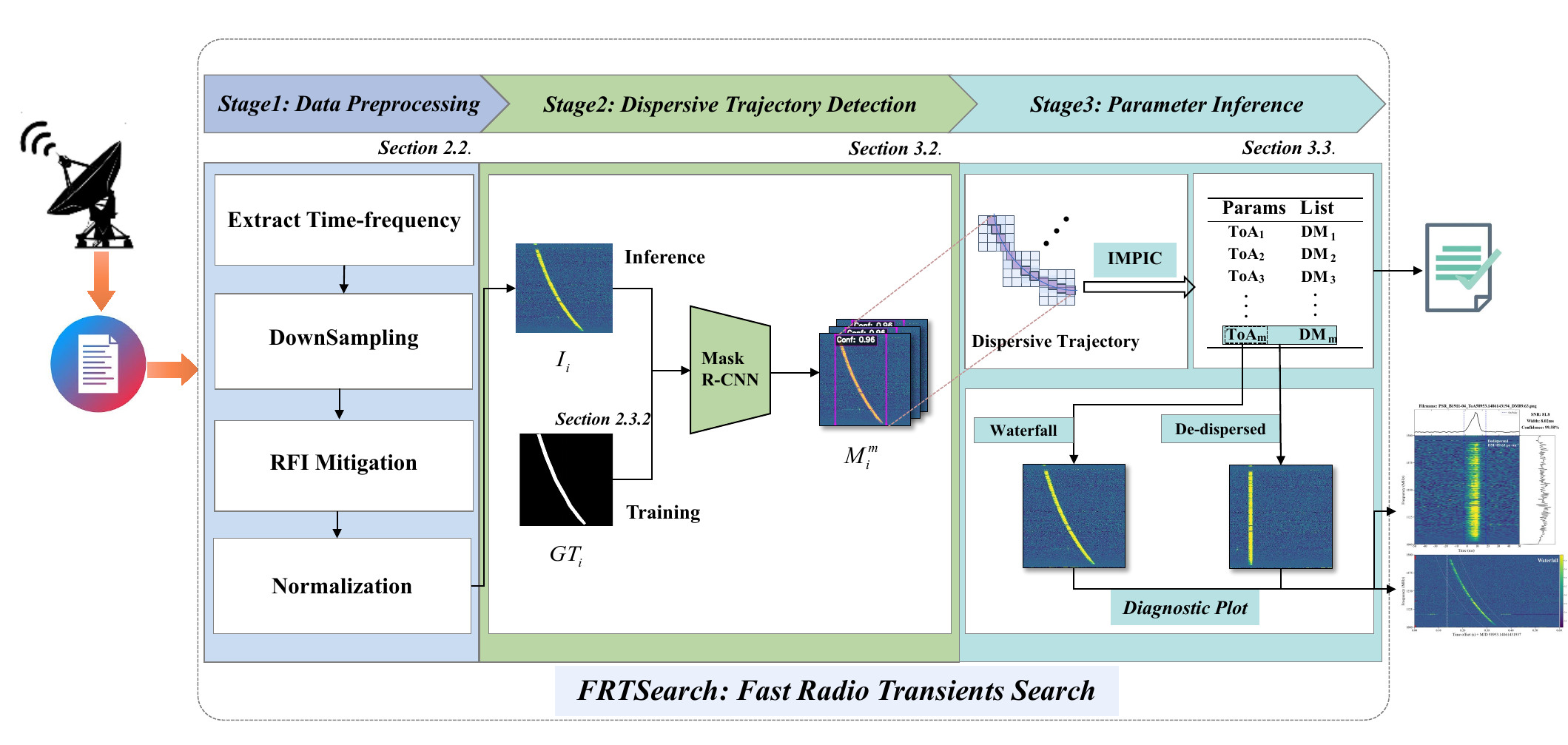}
    \caption{The end-to-end FRTSearch pipeline automatically processes radio observation files, generating diagnostic plots and a list of associated parameters for FRT candidates.
    }
    \label{fig:pipeline}
\end{figure}

The \texttt{FRTSearch} pipeline is structured into three distinct stages, as illustrated in Figure~\ref{fig:pipeline}: 
(i) Data Preprocessing (Section~\ref{sec:preprocessing}): Transforms raw radio observation files into standardized time-frequency dynamic spectra;  
(ii) Dispersive Trajectory Detection (Section~\ref{sec:maskrcnn}): Employs a Mask R-CNN architecture to localize and segment dispersive trajectories at the pixel level; and 
(iii) Parameter Inference (Section~\ref{sec:impec}): Applies the IMPIC algorithm to extract physical parameters ($\mathrm{DM}$ and $\mathrm{ToA}$) from the segmented trajectory coordinates via robust least-squares fitting. 
The pipeline ultimately outputs a candidate parameter catalog and associated diagnostic plots, facilitating automated verification and downstream scientific analysis.

\subsection{Dispersive Trajectory Detection via Mask R-CNN}
\label{sec:maskrcnn}

We employ the Mask R-CNN framework \citep{he2017mask}, implemented via the MMDetection toolbox \citep{chen2019mmdetection}, to perform instance segmentation on preprocessed dynamic spectra. Adopting the strategy used by PRESTO \citep{ransom2001}, we process individual observation files as independent units. In this work, each preprocessed CRAFTS file yields a single-channel image with dimensions $1 \times 256 \times 8{,}192$ (channel $\times$ frequency bins $\times$ time samples).

\begin{figure}[htbp]
    \centering
    \includegraphics[width=1.0\textwidth]{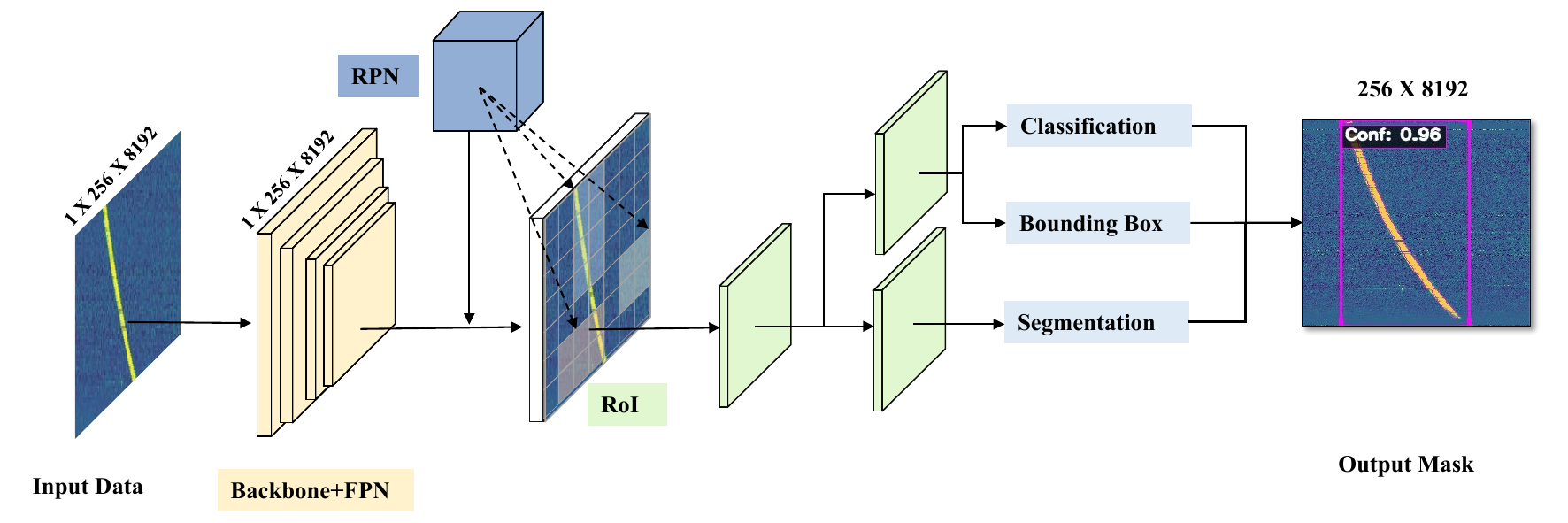}
    \caption{
        The architecture of the Mask R-CNN model designed for FRT detection. 
    }
    \label{fig:maskrcnn}
\end{figure}

\subsubsection{Mask R-CNN Architecture}

Formally, the detection process maps an input dynamic spectrum $I_i$ to a set of instance predictions:

\begin{equation}
\mathcal{M}_i = \text{Mask R-CNN}(I_i \mid \Theta)
\end{equation}
where $\Theta$ denotes the learned model parameters, and $\mathcal{M}_i = \{\mathcal{M}_i^1, \mathcal{M}_i^2, \ldots, \mathcal{M}_i^{N_i}\}$ represents the set of $N_i$ detected binary segmentation masks ($N_i \ge 0$). For the $j$-th detected instance, the pixel-level trajectory boundary is delineated by the mask $\mathcal{M}_i^j \in \{0,1\}^{256 \times 8192}$. Alongside each mask, the model also outputs a confidence score $s_i^j \in [0,1]$ indicating detection certainty, and a bounding box localizing the trajectory (see Figure~\ref{fig:pipeline_overview}, row 3). The extracted mask set $\mathcal{M}_i$ is subsequently passed to the IMPIC algorithm for parameter inference.

As illustrated in Figure~\ref{fig:maskrcnn}, the Mask R-CNN architecture processes each dynamic spectrum through a multi-stage pipeline. A backbone network, coupled with a Feature Pyramid Network (FPN; \citealt{lin2017}), extracts multi-scale features. These features are fed into a Region Proposal Network (RPN; \citealt{ren2015}), which generates candidate Regions of Interest (RoIs) potentially containing FRT signals. For each RoI, the network employs three parallel prediction heads: (i) a classification head that assigns FRT labels and confidence scores; (ii) a bounding-box regression head that refines object localization; and (iii) a mask segmentation head that predicts pixel-level masks delineating the dispersive trajectories.

The model is trained end-to-end by minimizing a multi-task objective function:
\begin{equation}
    \mathcal{L}_{\text{total}} = \mathcal{L}_{\text{cls}} + \mathcal{L}_{\text{box}} + \mathcal{L}_{\text{mask}}
\end{equation}
We employ Focal Loss ($\mathcal{L}_{\text{cls}}$;~\citealt{lin2017focal}), Smooth L1 Loss ($\mathcal{L}_{\text{box}}$;~\citealt{girshick2015fast}), and Dice Loss ($\mathcal{L}_{\text{mask}}$;~\citealt{milletari2016}) to optimize classification, bounding-box localization, and pixel-level segmentation, respectively.

\subsubsection{Performance Evaluation Metrics}
\label{sec:performance}

\textbf{Instance Segmentation Metrics.} Following standard computer vision benchmarks \citep{he2017mask}, we compute Average Precision (AP) at an Intersection over Union (IoU) threshold of 0.5:
\begin{equation}
\mathrm{IoU} = \frac{\mathrm{Area}(\mathrm{Prediction} \cap \mathrm{GT})}{\mathrm{Area}(\mathrm{Prediction} \cup \mathrm{GT})}
\end{equation}
We report both bounding-box AP (bbox AP), which measures localization accuracy, and segmentation AP (segm AP), which measures the quality of pixel-level trajectory delineation.

\textbf{Search Effectiveness Metrics.} To evaluate practical detection performance, we adopt lenient matching criteria (IoU $> 0.1$, confidence $> 0.1$) to maximize the capture of weak signals. Since observation files typically contain multiple FRT instances (Table~\ref{tab:dataset_summary}), we report Recall and FPPI (False Positives Per Input; \citealt{dollar2011}) at two distinct levels\footnote{True Positives (TP): predictions with confidence $> 0.1$ and IoU $> 0.1$ with any GT; False Positives (FP): predictions with confidence $> 0.1$ but IoU $\le 0.1$ to all GTs; False Negatives (FN): GT instances unmatched by any qualifying prediction.}:

\begin{itemize}
    \item \textbf{GT-level:} Assesses the detection completeness of discrete FRT instances. Recall $= \mathrm{TP}/(\mathrm{TP}+\mathrm{FN})$ quantifies the fraction of total GT instances successfully detected. FPPI $= \mathrm{FP}/N_{\mathrm{GT}}$ quantifies the number of false positives relative to the total count of GT instances.
    
    \item \textbf{File-level:} Evaluates the successful flagging of observation files. Recall measures the fraction of observation files containing at least one correctly detected FRT. FPPI $= \mathrm{FP}/N_{\text{files}}$ quantifies the average number of false positives per file.
\end{itemize}

File-level metrics are subsequently used in Section~\ref{sec:results_and_Analysis} to benchmark \texttt{FRTSearch} against traditional single-pulse search methods.

\subsubsection{Backbone Architecture Comparison}
\label{sec:backbone_comparison}

Dispersive trajectories exhibit unique morphological characteristics that pose significant challenges for feature extraction: they typically span the full frequency range (256 channels) with narrow temporal widths, creating extreme aspect ratios ($\sim$1:32 in $256 \times 8{,}192$ images; Figure~\ref{fig:pipeline_overview}). To identify the optimal backbone for capturing such specialized patterns, we systematically evaluated Mask R-CNN with seven backbone architectures: ResNet-18/50/101 \citep{he2016}, ResNeXt-101 \citep{xie2017aggregated}, HRNet \citep{sun2019deep}, Swin Transformer (Swin-T) \citep{liu2021swin}, and ConvNeXt \citep{liu2022convnet}.

All models were trained on a single NVIDIA GeForce RTX 3090 GPU (24 GB memory) for 36 epochs using identical data augmentation and learning rate (LR) schedules. To ensure optimal convergence, the optimizer configurations and base learning rates were tailored to align with the default implementations in the MMDetection framework \citep{chen2019mmdetection} (summarized in Table~\ref{tab:optimizer_config}). Specifically, standard CNN-based backbones (ResNet, ResNeXt, HRNet) utilized Stochastic Gradient Descent (SGD) \citep{bottou2010large}, whereas the Transformer-based Swin-T and modern ConvNeXt architectures employed AdamW optimization \citep{loshchilov2018decoupled} with specific parameter-wise weight decay strategies. The shared LR decay schedule included a linear warmup (1,000 iterations, starting factor of $10^{-3}$) followed by step decay ($\gamma = 0.1$) at epochs 27 and 33. Data augmentation strategies comprised: (i) random cropping to $256 \times 2{,}048$ patches to introduce spatial variability; and (ii) Copy-Paste augmentation \citep{ghiasi2021simple} with a probability of 0.4 (maximum 3 instances per image) to enrich instance diversity. The batch size was set to 2 due to memory constraints imposed by the large image dimensions.

\begin{table}[htbp]
\centering
\small
\caption{Optimization configurations for different backbone architectures.}
\label{tab:optimizer_config}
\begin{tabular}{lcccc}
\toprule
\textbf{Backbone} & \textbf{Optimizer} & \textbf{Base LR} & \textbf{Weight Decay} & \textbf{Specific Settings} \\
\midrule
ResNet/ResNeXt/HRNet & SGD & $0.02$ & $10^{-4}$ & Momentum = $0.9$ \\
Swin-T & AdamW & $10^{-4}$ & $0.05$ & $\beta=(0.9, 0.999)$, zero decay for pos\_embed/norm/bias \\
ConvNeXt & AdamW & $10^{-4}$ & $0.05$ & $\beta=(0.9, 0.999)$, layer-wise decay rate = $0.95$ \\
\bottomrule
\end{tabular}
\end{table}

Table~\ref{tab:maskrcnn_backbone_comparison} demonstrates that all evaluated backbones deliver robust detection performance, yielding segmentation metrics (bbox AP: 71.7--77.9\%; segm AP: 49.6--57.3\%) comparable to standard COCO benchmarks \citep{he2017mask}. All models maintain File Recall $\ge 96\%$, with ResNet-18, HRNet, Swin-T, and ConvNeXt successfully flagging 100\% of observation files containing FRTs. The high GT-level recall rates (92.7--99.4\%) further confirm that the models have effectively generalized the characteristic morphology required to robustly detect dispersive trajectories.

Although modern Swin-T and ConvNeXt architectures achieve exceptionally high GT-level recall ($\ge$98.1\%), their powerful representational capacity over-segments noisy dynamic spectra, resulting in severe false positive rates (File FPPI $\ge$ 34.1) and thousands of spurious candidates. A similar trade-off is observed with ResNet-18: its marginal 2.3\% recall improvement over HRNet incurs a doubled false alarm rate (File FPPI = 9.0 vs. 4.7), substantially increasing the downstream verification burden.

In contrast, HRNet demonstrates an optimal balance between detection completeness and efficiency. Unlike standard bottleneck architectures, HRNet \citep{sun2019deep} maintains high-resolution representations throughout the network, enabling precise delineation of narrow dispersive trajectories without amplifying background noise. It maintains competitive recall (95.1\% GT-level; 100\% File-level) while yielding the lowest false positive rates (GT FPPI = 0.8; File FPPI = 4.7) and restricting the total candidate count to only 1,357. Furthermore, HRNet achieves the highest segmentation fidelity (segm AP = 57.3\%), which is critical for the subsequent IMPIC algorithm that relies on precise pixel-level trajectory coordinates to invert the dispersion relation. We therefore selected HRNet for \texttt{FRTSearch} to guarantee the high-fidelity segmentation strictly required by the downstream IMPIC algorithm, while maintaining robust false positive mitigation in noisy dynamic spectra.

\begin{table*}[htbp]
\centering
\small
\caption{Backbone architecture comparison on the validation set (100 files, 587 GT instances). The selected model is marked with $\star$.}
\label{tab:maskrcnn_backbone_comparison}
\begin{tabular*}{\textwidth}{@{\extracolsep{\fill}}lcccccccc}
\toprule
\multirow{2}{*}{\textbf{Backbone}} 
& \multirow{2}{*}{\textbf{TP}} 
& \multirow{2}{*}{\textbf{\# Candidates}} 
& \multicolumn{2}{c}{\textbf{Segmentation Metrics}} 
& \multicolumn{2}{c}{\textbf{GT-level Metrics}} 
& \multicolumn{2}{c}{\textbf{File-level Metrics}} \\
\cmidrule(lr){4-5} \cmidrule(lr){6-7} \cmidrule(lr){8-9}
& & & \textbf{bbox AP (\%)} & \textbf{segm AP (\%)} 
& \textbf{Recall (\%)} & \textbf{FPPI} 
& \textbf{Recall (\%)} & \textbf{FPPI} \\
\midrule
ResNet-18        & 1237 & 2143          & 75.8          & 52.7          & 97.4 & 1.5         & \textbf{100.0} & 9.0 \\
ResNet-50        & 1001          & 1839          & 72.9          & 49.6          & 92.7          & 1.4        & 96.0           & 8.3  \\
ResNet-101       & 1058          & 3046          & 72.9          & 52.3          & 95.9          & 3.3          & 97.0           & 19.8 \\
ResNeXt-101      & 990           & 2395          & 74.7          & 51.0          & 95.2          & 2.3         & 97.0           & 14.0 \\
HRNet$\star$     & 887           & \textbf{1357} & \textbf{77.9} & \textbf{57.3} & 95.1          & \textbf{0.8} & \textbf{100.0} & \textbf{4.7} \\
Swin-T   & 1278          & 4688          & 71.7          & 51.6          & 98.1          & 5.8          & \textbf{100.0} & 34.1 \\
ConvNeXt  & \textbf{1327}           & 5982 & 75.3 & 56.0 & \textbf{99.4}          & 7.9 & \textbf{100.0} & 46.5 \\
\bottomrule
\end{tabular*}
\end{table*}

\subsection{Parameter Inference via IMPIC Algorithm}
\label{sec:impec}

Following the detection and segmentation of dispersive trajectories by the Mask R-CNN model, the IMPIC algorithm extracts physical parameters by mapping the segmented binary masks to time-frequency coordinates. For each detected instance mask $\mathcal{M}_i = \{\mathcal{M}_i^j\}$, we extract the time-frequency coordinate set $\mathcal{P}_j = \{(\nu_k,t_k ) \mid \mathcal{M}_i^j[\nu_k, t_k] = 1\}$ representing the dispersive trajectory. The DM is then inferred through robust fitting to the cold plasma dispersion relation \citep{lorimer2005}:

\begin{equation}
\mathrm{DM} = \frac{\Delta t}{k_{\mathrm{DM}} \times \Delta(\nu^{-2})}, \quad \text{where} \quad \Delta(\nu^{-2}) = \nu_{\mathrm{low}}^{-2} - \nu_{\mathrm{high}}^{-2}
\label{eq:dm}
\end{equation}

To mitigate the impact of segmentation outliers (e.g., mask boundary irregularities or intersecting trajectories from overlapping bursts), IMPIC employs a Random Sample Consensus (RANSAC) fitting strategy \citep{fischler1981random}. The algorithm iteratively samples subsets of trajectory points to perform Nonlinear Least-Squares (NLS) optimization \citep{virtanen2020scipy}. This optimization minimizes the residuals between the segmented pixel coordinates and the theoretical dispersion curve defined by Equation~\eqref{eq:dispersion}, ultimately selecting the candidate solution that maximizes the coefficient of determination ($R^2$). The ToA is subsequently derived by: (i) identifying the earliest arrival time $(t_{\mathrm{earliest}}, \nu_{\mathrm{earliest}})$ from the segmented trajectory; and (ii) extrapolating to the observational high frequency $\nu_{\mathrm{high}}$ via Equation~\eqref{eq:dispersion}. The complete workflow is detailed in Algorithm~\ref{algo:IMPIC}.

\begin{algorithm}[htbp]
\caption{\texttt{IMPIC} - Iterative Mask-based Parameter Inference and Calibration}
\label{algo:IMPIC}
\begin{algorithmic}
\Require Detected instances \textbf{$\mathcal{M}_i = \{\mathcal{M}_i^1, \ldots, \mathcal{M}_i^{N_i}\}$}, iterations $N_{\rm iter}$, sample size $N_{\rm sample}$, high frequency $\nu_{\rm high}$
\Ensure Parameter matrix $T_{\rm out} \in \mathbb{R}^{N_i \times 2}$ containing [DM, ToA] for each instance
\State Initialize $T_{\rm out} \leftarrow$ empty matrix
\For{$j = 1, 2, \ldots, N_i$}
    \State Extract mask coordinates: $\mathcal{P}_j \leftarrow \{(t_k, \nu_k) \mid \mathcal{M}_i^j[t_k, \nu_k] = 1\}$
    \If{$|\mathcal{P}_j| < 3$}
        \State $T_{\rm out}[j] \leftarrow [0, 0]$ and \textbf{continue}
    \EndIf
    \State ${\rm best\_score} \leftarrow -\infty$, $\widehat{\rm DM} \leftarrow 0$
    \For{$n = 1$ \textbf{to} $N_{\rm iter}$}
        \State Randomly sample $\mathcal{S} \subset \mathcal{P}_j$ with $|\mathcal{S}| = N_{\rm sample}$
        \State Fit DM via NLS on Eq.~\eqref{eq:dm}
        \State Compute $R^2$ score for current fit
        \If{$R^2 > {\rm best\_score}$}
            \State ${\rm best\_score} \leftarrow R^2$, $\widehat{\rm DM} \leftarrow$ fitted DM
        \EndIf
    \EndFor
    \State Find earliest arrival: $(t_{\rm earliest}, \nu_{\rm earliest}) \leftarrow \arg\min_{(t,\nu) \in \mathcal{P}_j} t$
    \State Compute time offset: $\Delta t \leftarrow k_{\rm DM} \times \widehat{\rm DM} \times (\nu_{\rm earliest}^{-2} - \nu_{\rm high}^{-2})$ \Comment{Eq.~\eqref{eq:dispersion}}
    \State $\widehat{\rm ToA} \leftarrow t_{\rm earliest} - \Delta t$ \Comment{ToA at $\nu_{\rm high}$}
    \State $T_{\rm out}[j] \leftarrow [\widehat{\rm DM}, \widehat{\rm ToA}]$
\EndFor
\State \Return $T_{\rm out}$
\end{algorithmic}
\end{algorithm}

\subsubsection{IMPIC Performance Evaluation and Error Analysis}
\label{sec:error_analysis}

\begin{figure*}[htbp]
    \centering
    \includegraphics[width=0.70\textwidth]{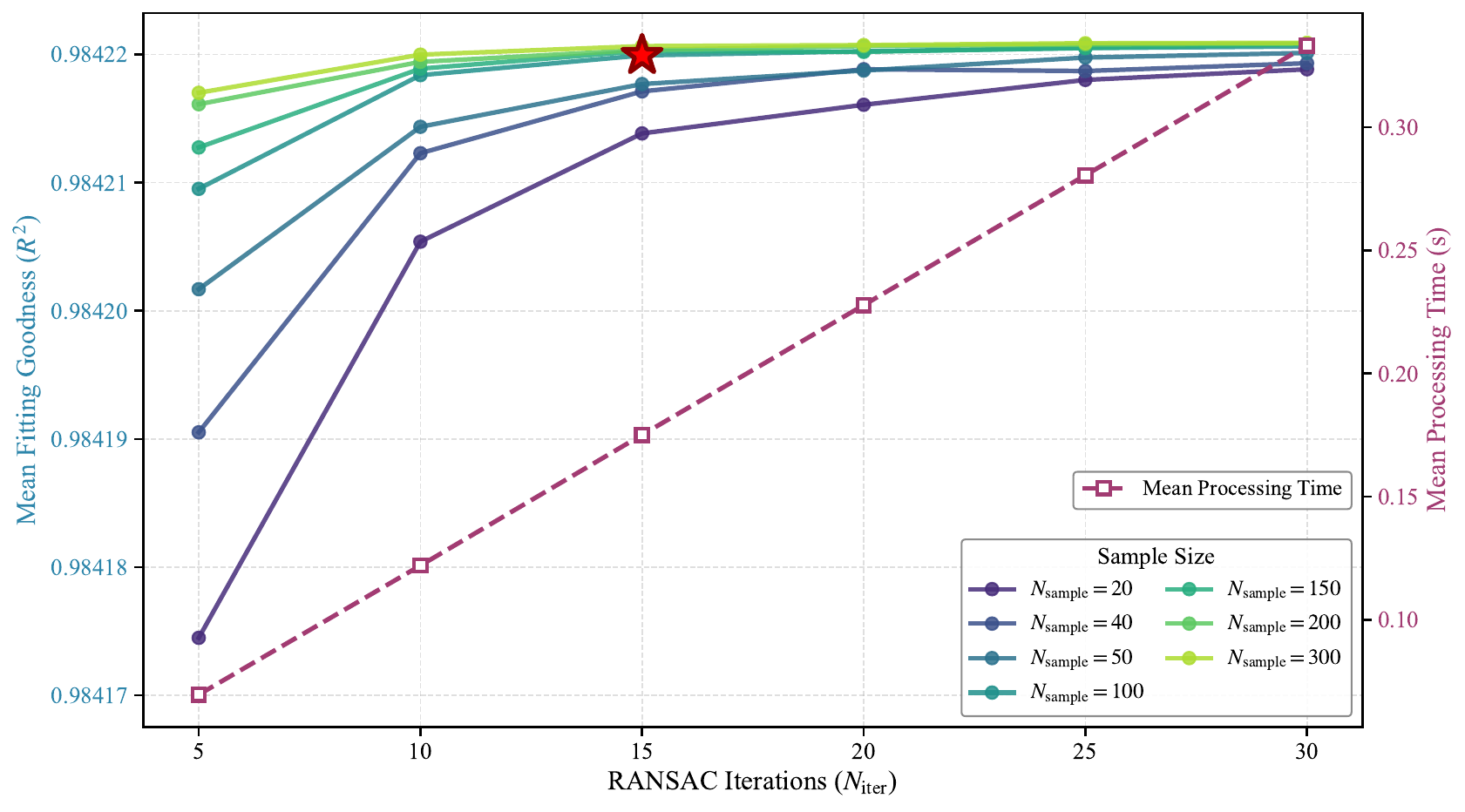}
    \caption{Evaluation of IMPIC hyperparameters. The plot compares the mean $R^2$ score (left axis, solid lines) and processing time (right axis, dashed line) across different sample sizes ($N_{\mathrm{sample}}$) and iterations ($N_{\mathrm{iter}}$). The red star highlights the adopted configuration ($N_{\mathrm{sample}}=100$, $N_{\mathrm{iter}}=15$) used in this work.}
    \label{fig:hyperparameter_opt}
\end{figure*}

Based on a systematic grid search (Figure~\ref{fig:hyperparameter_opt}), we configured IMPIC with $N_{\mathrm{sample}}=100$ trajectory points per RANSAC iteration and $N_{\mathrm{iter}}=15$ iterations. This configuration optimally balances computational efficiency (average inference time $\sim$0.18~s) with robust fitting capabilities. We evaluated the algorithm on 887 true positive detections from the HRNet-based Mask R-CNN validation set, comprising 743 pulsars, 3 RRATs, and 141 FRBs with DMs ranging from 14.9 to 565~pc~cm$^{-3}$ (Table~\ref{tab:dm_accuracy}). The algorithm achieves weighted average accuracies of 83.6\% and 96.0\% within the tolerances of $1\sigma$ and $2\sigma$, respectively, demonstrating robust performance under various 
observational conditions.

IMPIC maintains a high goodness of fit ($R^2 \ge 0.96$) across all sources (Table~\ref{tab:dm_accuracy}), indicating that the segmented trajectories faithfully conform to the dispersion relation (Equation~\ref{eq:dispersion}). This geometric fidelity empirically validates the core premise of our approach: that physical parameters can be accurately inferred via direct inversion of pixel-level trajectory coordinates. Quantitatively, the inferred DM values show robust consistency with catalog references, exhibiting a weighted mean systematic offset of $+3.4$~pc~cm$^{-3}$ and a weighted fractional error of $3.8\%$ across all validation sources. This performance confirms that IMPIC provides sufficiently accurate priors for downstream signal validation. The physical validity of these parameters is further corroborated by the coherent pulse profiles recovered in the de-dispersed dynamic spectra using IMPIC-inferred DM values (Figure~\ref{fig:pipeline_overview}, row 4). The sharp temporal concentration of signal power confirms that the algorithm successfully reconstructs the intrinsic emission timescale, demonstrating end-to-end consistency from trajectory segmentation to physical characterization.

\begin{table*}[htbp]
\centering
\small
\caption{DM Inference Statistics for Validation Sources. Columns: reference DM; number of true positive detections (TP); mean and standard deviation of inferred DMs; mean coefficient of determination ($R^2$) of the trajectory fit; and fractional accuracy within $\pm1\sigma$ and $\pm2\sigma$ tolerance.}
\label{tab:dm_accuracy}
\begin{tabular*}{\textwidth}{@{\extracolsep{\fill}}lccccccc}
\toprule
\textbf{Source} & \textbf{Ref DM (pc~cm$^{-3}$)} & \textbf{TP} & \textbf{Mean DM (pc~cm$^{-3}$)} & \textbf{Std (pc~cm$^{-3}$)} & \boldmath$  \mathbf{R^2}$ & \boldmath$\mathbf{1\sigma}$ \textbf{(\%)} & \boldmath$\mathbf{2\sigma}$ \textbf{(\%)} \\
\midrule
RRAT J2325-0530 & 14.9 & 3 & 19.7 & 2.1 & 0.969 & 66.7 & 100.0 \\
PSR B0820+02 & 23.7 & 325 & 30.0 & 14.9 & 0.965 & 92.6 & 95.1 \\
PSR B1911-04 & 89.4 & 103 & 88.3 & 3.9 & 0.991 & 90.3 & 97.1 \\
PSR J1820-0509 & 102.4 & 109 & 101.5 & 4.0 & 0.995 & 89.9 & 96.3 \\
PSR J0652-0142 & 116.3 & 42 & 113.8 & 6.2 & 0.992 & 73.8 & 95.2 \\
PSR J1948+2333 & 198.2 & 72 & 200.1 & 14.4 & 0.995 & 90.3 & 95.8 \\
PSR J1845-0545 & 315.9 & 92 & 321.7 & 13.1 & 0.996 & 87.0 & 95.7 \\
FRB 20201124A & 420.0 & 77 & 411.7 & 25.1 & 0.985 & 76.6 & 93.5 \\
FRB 20121102 & 565.0 & 64 & 568.1 & 32.0 & 0.994 & 85.9 & 95.3 \\
\bottomrule
\end{tabular*}
\end{table*}

\begin{figure*}[htbp]
    \centering
    \includegraphics[width=0.95\textwidth]{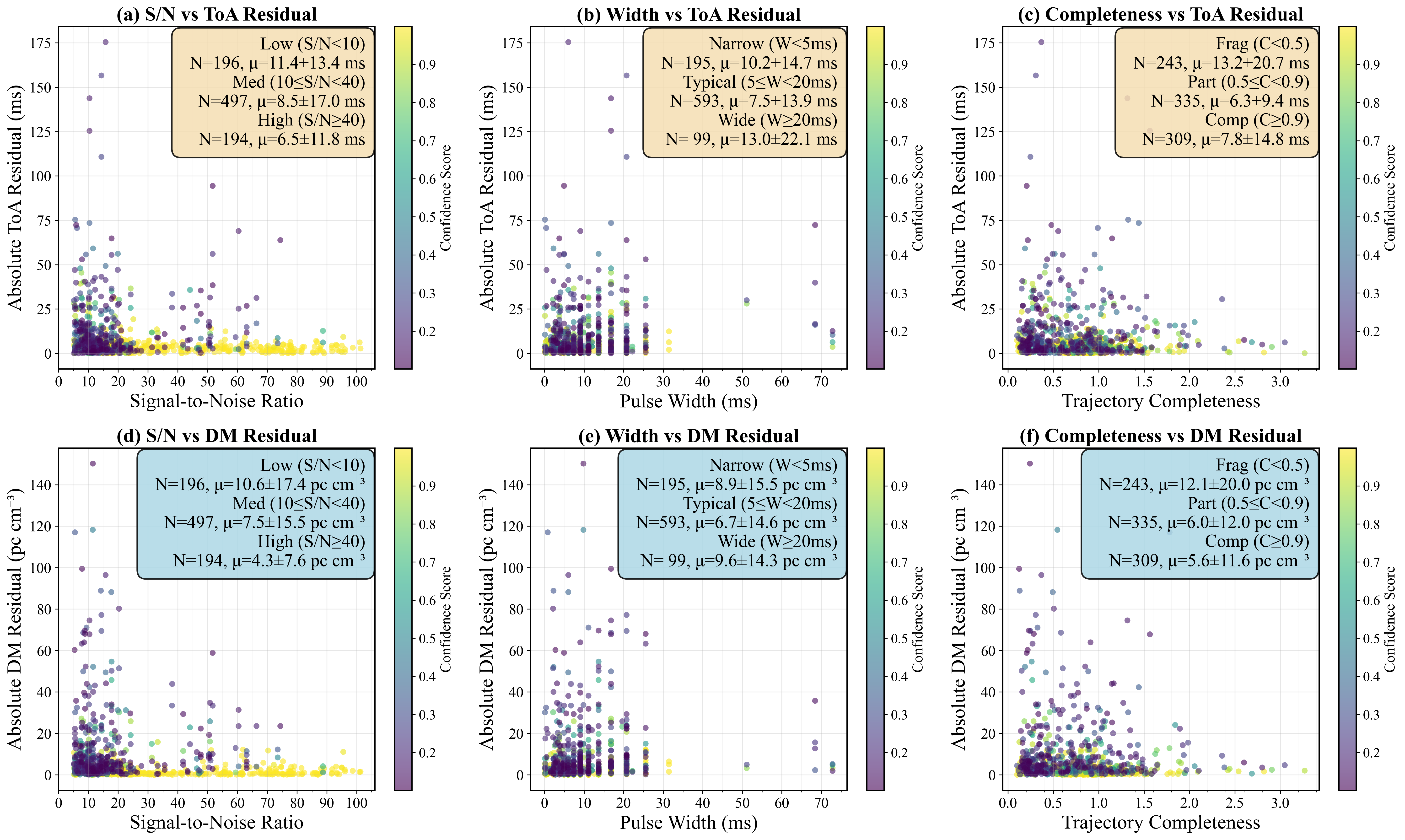}
    \caption{Error correlation with observational parameters. Scatter plots show absolute ToA residuals (top row: a--c) and DM residuals (bottom row: d--f) versus S/N (left column), pulse width (middle column), and trajectory completeness (right column). Points are color-coded by detection confidence scores.  }
    \label{fig:error_factors}
\end{figure*}

To quantify parameter inference uncertainties, we systematically analyzed the correlation between residuals and observational parameters: S/N, pulse width, and trajectory completeness ($\mathcal{C} = |\mathcal{P}_{\mathrm{pred}}|/|\mathcal{P}_{\mathrm{GT}}|$, where $\mathcal{P}$ denotes the time-frequency coordinate set extracted from segmentation masks) (Figure~\ref{fig:error_factors}).

Inference precision exhibits a strong inverse correlation with S/N (panels a, d). As S/N increases from $<10$ to $\ge 40$, mean absolute errors decrease significantly by 42.8\% for ToA (from $11.4 \pm 13.4$~ms to $6.5 \pm 11.8$~ms) and by 59.1\% for DM (from $10.6 \pm 17.4$~pc~cm$^{-3}$ to $4.3 \pm 7.6$~pc~cm$^{-3}$). This trend reflects the inherent challenge of segmenting faint trajectories in noise-dominated backgrounds: mask boundary uncertainties from the Mask R-CNN stage directly propagate to coordinate extraction errors in $\mathcal{P}_{\mathrm{pred}}$, subsequently degrading RANSAC fitting accuracy in the IMPIC algorithm. Notably, predictions cluster at elevated S/N, validating the model's intrinsic uncertainty quantification.

Similarly, inference stability is critically governed by trajectory completeness (panels c, f). Fragmented trajectories ($\mathcal{C} < 0.5$) yield elevated residuals ($12.1 \pm 20.0$~pc~cm$^{-3}$) due to insufficient geometric constraints for robust fitting of the dispersion relation (Equation~\ref{eq:dm}). Conversely, complete trajectories ($\mathcal{C} \ge 0.9$) provide sufficient leverage across the frequency band, reducing DM errors by 54.2\% (to $5.6 \pm 11.6$~pc~cm$^{-3}$) and ToA errors by 40.7\%. This underscores the critical importance of high-fidelity segmentation for precise parameter estimation. Furthermore, residuals remain largely uniform across pulse widths of 0.1--70~ms (panels b, e), confirming that IMPIC inference relies exclusively on the geometric curvature of the dispersive sweep (Equation~\ref{eq:dispersion}) rather than temporal profile characteristics.

\subsection{Diagnostic Plot Generation}

\begin{figure}[htbp]
    \centering
    \includegraphics[width=1.0\textwidth]{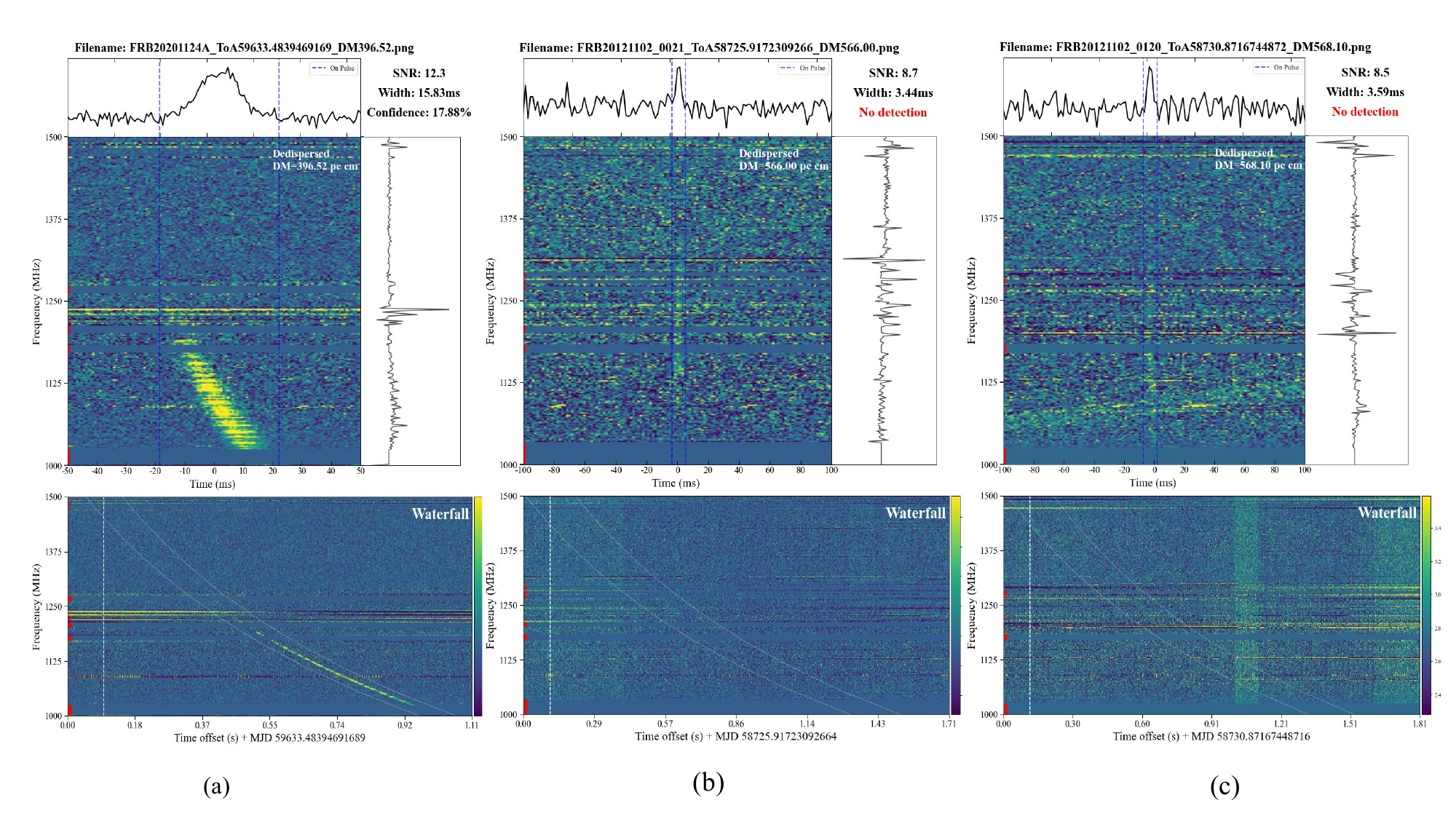}
    \caption{Diagnostic plots for three representative candidates. Each column displays: (top) de-dispersed pulse profile with on-pulse region (vertical dotted lines) and metadata; (middle) de-dispersed dynamic spectrum with frequency-averaged bandpass (right); (bottom) dispersed dynamic spectrum showing frequency-time trajectory. (a) Candidate using IMPIC-inferred parameters. (b--c) FRB 20121102 (missed by \texttt{FRTSearch}) using parameters provided by FAST-FREX \citep{guo2024}.
    }
    \label{fig:candidate_b1911-04} 
\end{figure}

Leveraging the inferred ToA and DM parameters, \texttt{FRTSearch} automatically synthesizes standard diagnostic plots---comprising waterfall plots, de-dispersed dynamic spectra, and pulse profiles---to facilitate manual verification. Figure~\ref{fig:candidate_b1911-04} presents representative examples generated by our pipeline. Panel (a) illustrates a detection of FRB~20201124A with moderate deviations ($\Delta t \approx 25$~ms, $\Delta\mathrm{DM} \approx -23$~pc~cm$^{-3}$), resulting in imperfect de-dispersion. Although this deviation manifests as visible residual curvature in the de-dispersed spectrum, the astrophysical signal remains prominently centered and visually distinct. 
This visual robustness empirically confirms that \texttt{FRTSearch}'s inference precision (weighted fractional error of $3.8\%$ across validation sources) is sufficient for effective candidate validation. For scientific applications necessitating high-precision timing, these inferred parameters serve as reliable priors for efficient refinement via localized grid searches.
Furthermore, panels (b) and (c) present examples of missed detections under extreme RFI; a detailed morphological analysis of these failure cases is deferred to Section~\ref{subsec:sensitivity}.

\section{Experiments and Results Analysis}
\label{sec:experiments}

\subsection{Benchmark Comparison with Single-Pulse Search Methods}
\label{sec:fast_frb_dataset}

We evaluated the performance of \texttt{FRTSearch} against established single-pulse search tools using the publicly available FAST Dataset for Fast Radio bursts EXploration (FAST-FREX; \citealt{guo2024}) as a benchmark. This dataset comprises 600 verified burst samples originating from three distinct repeating FRBs. Each sample is stored as an individual FITS file with a duration of approximately 6.04~s. All methods performed blind searches across the full parameter space without prior knowledge, ensuring a fair evaluation of their capability to discover unknown transients under real-world systematic survey conditions.

\subsubsection{Experimental Setup}

To ensure a rigorous comparison, all single-pulse search pipelines processed an identical DM trial grid. De-dispersion was performed across a DM range of 3.0 to 1{,}200~pc~cm$^{-3}$ using a variable spacing step increasing from 0.1 to 1.0~pc~cm$^{-3}$, yielding a total of 3{,}975 trial DM values. We compared \texttt{FRTSearch} against three widely adopted pipelines:

\begin{itemize}
    \item \textbf{PRESTO} \citep{ransom2001}: A GPU-accelerated pipeline \citep{you2021} incorporating RFI mitigation (\texttt{rfifind}), de-dispersion (\texttt{prepsubband}), and single-pulse searching (\texttt{single\_pulse\_search.py}). Candidates with S/N $\geq 5$ were retained following default settings, with diagnostic plots generated for verification.
    
    \item \textbf{Heimdall} \citep{barsdell2012}: A GPU-accelerated pipeline executed with default configurations and an S/N threshold $\geq 5$. This method outputs parameter lists only, without generating diagnostic plots.
    
    \item \textbf{TransientX} \citep{men2024}: Executed with 4 MPI parallel processes and an S/N threshold $\geq 5$. All other parameters were maintained at default values. This method generates diagnostic plots.
\end{itemize}

Unlike single-pulse search methods that rely on S/N thresholds, \texttt{FRTSearch} evaluates the probability of a signal being an FRT based on a confidence score. As illustrated in Figure~\ref{fig:error_factors}, many low-S/N FRTs exhibit relatively low confidence scores (approximately 0.1). Consequently, we configured \texttt{FRTSearch} with a conservative confidence threshold of 0.1 to optimally balance detection sensitivity with false positive suppression. The pipeline generates parameters and diagnostic plots for all candidates exceeding this threshold.

Furthermore, to enable fair speed comparisons against methods with differing output formats, we measured \texttt{FRTSearch} execution time under two operational modes: parameter-only output (comparable to Heimdall) and full pipeline with diagnostic plot generation (comparable to PRESTO and TransientX). All experiments were conducted on a server equipped with an Intel Xeon Platinum 8255C CPU, an NVIDIA Tesla T4 GPU, and 32~GB of memory.

\begin{table*}[htbp]
\centering
\caption{Benchmark comparison on the FAST-FREX dataset \cite{guo2024} (600 FRB bursts). Best results are highlighted in \textbf{bold}. For \texttt{FRTSearch}, execution time is reported as \textit{parameter-only / with diagnostic plots}.}
\label{tab:comparison_fast_frex}
\begin{tabular*}{\textwidth}{@{\extracolsep{\fill}}lcccccc}
\toprule
\textbf{Method} & \textbf{Threshold} & \textbf{TP} & \textbf{\#Candidates} & \textbf{Recall (\%)} & \textbf{FPPI (per file)} & \textbf{Time (s/file)} \\
\midrule
PRESTO          & S/N $\ge$ 5           & 479         & 2{,}990{,}577               & 79.8                 & 4{,}983.5           & 221.8 \\
Heimdall        & S/N $\ge$ 5           & 505         & 29{,}901             & 84.2                 & 49.0          & 16.0 \\
TransientX      & S/N $\ge$ 5           & \textbf{600}         & 49{,}405               & \textbf{100.0}               & 81.3            & 60.3 \\
\midrule
FRTSearch & Confidence $\ge$ 0.1 & 588 & \textbf{3{,}050}       & 98.0        &  \textbf{4.1}  & \textbf{8.7} / \textbf{16.0} \\
\bottomrule
\end{tabular*}
\end{table*}

\subsubsection{Results and Analysis} \label{sec:results_and_Analysis}

Table~\ref{tab:comparison_fast_frex} summarizes the benchmark results on the FAST-FREX dataset, comparing \texttt{FRTSearch} against widely adopted single-pulse search pipelines (PRESTO, Heimdall, and TransientX). These results empirically validate that our ``detect-and-infer'' 
paradigm effectively circumvents the efficiency-sensitivity trade-off inherent in traditional exhaustive search methods.

\texttt{FRTSearch} demonstrates detection capabilities competitive with the most exhaustive search approaches. At a confidence threshold of 0.1, our method achieves a recall of 98.0\% (588/600), significantly outperforming PRESTO (79.8\%) and Heimdall (85.0\%), while closely approaching TransientX (100\%). The 12 missed events primarily correspond to extremely faint bursts where the dispersive trajectory is barely distinguishable from background noise (see Figure~\ref{fig:candidate_b1911-04}c), representing the intrinsic sensitivity limit for image-based detection methods.

A key advantage of \texttt{FRTSearch} is its ability to drastically reduce false positives. As shown in Table~\ref{tab:comparison_fast_frex}, \texttt{FRTSearch} attains an FPPI of only 4.1 while maintaining a recall of 98.0\%. This corresponds to a false positive reduction of over 99.9\% relative to PRESTO (FPPI $\approx 4{,}983.5$) and approximately 92\%--95\% relative to Heimdall and TransientX. This high purity is achieved because \texttt{FRTSearch} enforces a robust morphological constraint. By training on the universal dispersive trajectory governed by the dispersion relation (Equation~\ref{eq:dispersion}), the pixel-level instance segmentation network effectively distinguishes genuine astrophysical signals from RFI, which typically lacks the strict coherent $\nu^{-2}$ dispersive trajectory governed by cold plasma physics. This capability significantly alleviates the manual verification workload, effectively producing a high-purity candidate set.

The significant improvements in processing efficiency stem from the architectural design of \texttt{FRTSearch}. By bypassing the exhaustive de-dispersion and matched filtering loops over dense DM and pulse width grids that characterize traditional pipelines, our method avoids redundant computation. In parameter-only mode, \texttt{FRTSearch} processes a standard 6.04~s FAST-FREX observation file in an average of 8.7~s (including $\sim$5.83~s for preprocessing, $\sim$0.52~s for Mask R-CNN detection, and $\sim$0.69~s for IMPIC inference, with the remainder attributed to model initialization and I/O operations), representing a $1.8\times$ speedup over Heimdall. When generating full diagnostic plots, the processing time is 16.0~s, achieving speedups of $3.8\times$ and $13.9\times$ over TransientX (60.3~s) and PRESTO (221.8~s), respectively.

In their evaluation on the FAST-FREX dataset, \citet{guo2024} reported an 83.5\% recall and 3.37~s/file for the RaSPDAMv2 object detection method by utilizing prior knowledge to restrict the search to a narrow DM range (350--650~pc~cm$^{-3}$). A direct quantitative comparison is precluded because all four methods in Table~\ref{tab:comparison_fast_frex} strictly perform blind searches across the full 3.0--1{,}200~pc~cm$^{-3}$ parameter space. Furthermore, as explicitly stated by \citet{guo2024}, their object detection approach cannot measure DM values. In contrast, \texttt{FRTSearch} advances beyond simple localization: by leveraging pixel-level instance segmentation, our IMPIC algorithm autonomously infers precise physical parameters (DM and ToA), realizing the ``detect-and-infer'' paradigm for the discovery and characterization of astrophysical sources.

\subsection{Cross-Facility Validation: ASKAP FRB Dataset}
\label{sec:askap_test}

To evaluate cross-observatory generalization, we applied the CRAFTS-FRT trained \texttt{FRTSearch} directly to 19 FRBs detected by the Australian Square Kilometre Array Pathfinder (ASKAP) \citep{shannon2018} without retraining. As summarized in Table~\ref{tab:askap_vs_fast}, ASKAP presents substantial instrumental differences from CRAFTS: 13-fold coarser temporal resolution (1.266~ms vs. 98.304~$\mu$s), 8-fold coarser spectral resolution (1~MHz vs. 0.122~MHz channels), different frequency coverage (1130--1470~MHz vs. 1000--1500~MHz), and distinct data formats. The test sample spans DM = 114--992~pc~cm$^{-3}$ and S/N = 8.2--34.7, providing a challenging cross-facility benchmark.

\begin{figure}[htbp]
    \centering
    \includegraphics[width=1.0\textwidth]{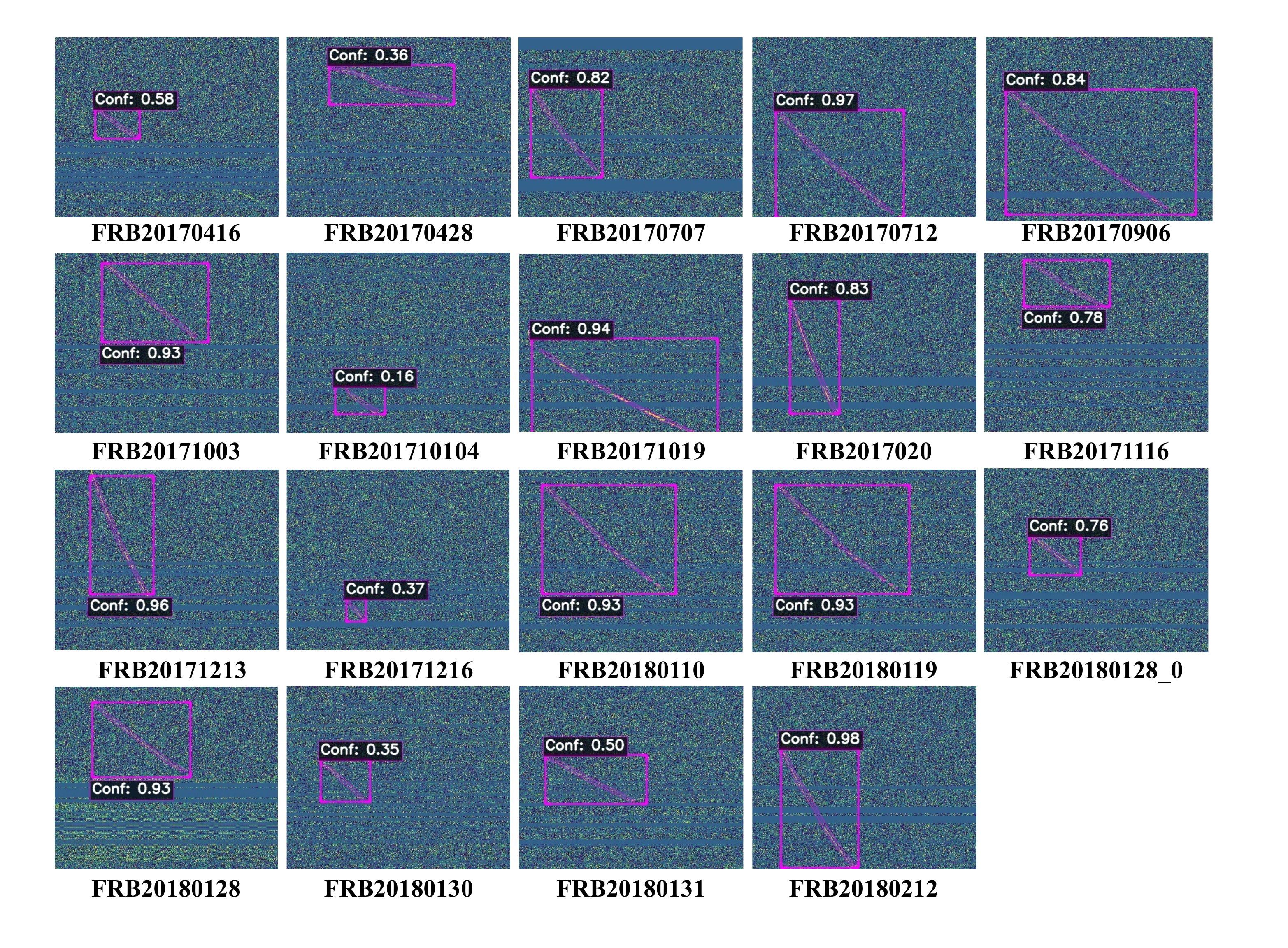}
    \caption{
        \texttt{FRTSearch} detection results on 19 ASKAP FRBs \citep{shannon2018}. The panels show time-frequency spectra ($336 \times 336$ pixels) with overlaid model predictions: segmentation masks (pink contours), bounding boxes (pink rectangles), and confidence scores (white text).
    }
    \label{fig:ska_results}
\end{figure}

Even without exposure to ASKAP data during training, \texttt{FRTSearch} achieved 100\% recall, successfully detecting all 19 FRBs. Figure~\ref{fig:ska_results} visualizes the Mask R-CNN detection results for these events, showing morphologically correct segmentation masks, while Table~\ref{tab:askap_detection_results} presents the complete detection summary, including DM and ToA values inferred via the IMPIC algorithm. Notably, the detection confidence scores exhibit significant variance (0.16--0.98). Consistent with the morphological dependencies analyzed in Section~\ref{sec:error_analysis}, this variance is driven by intrinsic burst characteristics rather than instrumental artifacts. Specifically, faint bursts affected by severe dispersion smearing manifest as fragmented, narrowband emissions (e.g., FRB 20171216). The classification network assigns lower confidence to these degraded trajectories because their fragmented nature provides significantly fewer discriminative pixel-level features for robust classification.

Despite this variance in confidence, the successful recovery of all events demonstrates that by exploiting the universal dispersive trajectories in time--frequency dynamic spectra, the model effectively identifies signals across disparate radio-astronomical datasets. Instead of overfitting to instrument-specific characteristics, \texttt{FRTSearch} focuses on intrinsic physical signatures, thereby facilitating the robust discovery of astrophysical sources across heterogeneous facilities.

\begin{table}[htbp]
    \centering
    \caption{Comparison of instrumental parameters between CRAFTS-FRT (training data) and ASKAP (test data).}
    \label{tab:askap_vs_fast}
    \begin{tabular}{lcc}
        \toprule
        \textbf{Parameter} & \textbf{CRAFTS} & \textbf{ASKAP} \\
        \midrule
        File Format & PSRFITS & Filterbank \\
        Sampling Time ($\mu$s) & 98.3 & 1266.5 \\
        Channel Width (MHz) & 0.122 & 1.0 \\
        Bit Depth  & 2 & 8 \\
        Integration Time (s) & 12.88 & 3345.33 \\
        Frequency Range (MHz) & 1000--1500 & 1130--1465 \\
        Number of Channels & 4096 & 336 \\
        Time-Freq.\ Downsampling & 16$\times$, 16$\times$ & 1$\times$, 1$\times$ \\
        \bottomrule
    \end{tabular}
\end{table}

\section{Discussion and Limitations}
\label{sec:limitations}

\subsection{Sensitivity Limits and Dispersion Smearing}
\label{subsec:sensitivity}

\begin{figure}[htbp]
    \centering
    \includegraphics[width=0.7\textwidth]{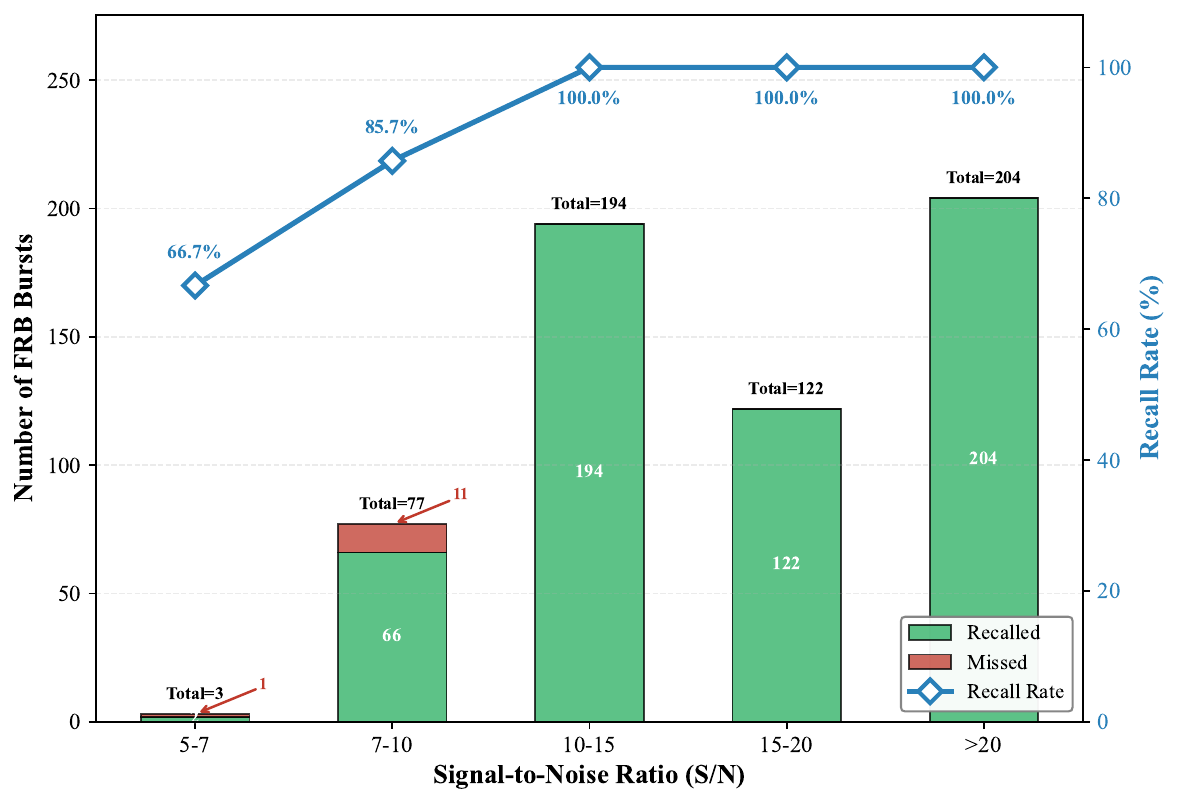}
    \caption{
        Detection efficiency of \texttt{FRTSearch} on the FAST-FREX benchmark. The recall rate is evaluated as a function of S/N, revealing a progressive decrease in detection reliability for $\mathrm{S/N} < 10$.
    }
    \label{fig:recall_vs_snr}
\end{figure}

Despite achieving a 98.0\% recall rate on the FAST-FREX benchmark, \texttt{FRTSearch} missed 12 out of 600 confirmed bursts. To quantitatively characterize the detection sensitivity, we evaluated the recall rate as a function of S/N measured using TransientX \citep{men2024} (Figure~\ref{fig:recall_vs_snr}). \texttt{FRTSearch} achieves 100\% recall (520/520) for bursts with $\mathrm{S/N} \geq 10$. However, the recall drops to 85.7\% (66/77) in the $7 \leq \mathrm{S/N} < 10$ range and further to 66.7\% (2/3) for $5 \leq \mathrm{S/N} < 7$. This reveals a progressive performance degradation for weaker signals, indicating a practical sensitivity threshold near $\mathrm{S/N} \approx 10$.

When interpreting this limitation, it should be noted that the reported S/N values are derived from traditional single-pulse search methods \textit{after} de-dispersion and matched filtering \citep{Cordes_2003}. These steps temporally integrate energy, rendering intrinsically faint bursts more prominent in the de-dispersed dynamic spectra (Figure~\ref{fig:candidate_b1911-04}b and c, middle row). In contrast, \texttt{FRTSearch} operates directly on the raw dispersed waterfall plots (bottom row). In this domain, dispersion smearing distributes the limited energy of faint bursts across time, rendering their trajectories highly diffuse and reducing their pixel-level features below the threshold required for successful segmentation.

A morphological analysis of the 12 missed bursts reveals that this smearing effect is often compounded by other observational challenges. As demonstrated by the two representative examples in Figure~\ref{fig:candidate_b1911-04}b and c, these bursts exhibit intrinsically narrowband emission, spanning only a limited segment of the observation band. Consequently, their already diluted and fragmented trajectories are further degraded by overlapping interference. Specifically, the burst in panel (b) is obscured by adjacent narrowband RFI at 1125--1300~MHz, while the burst in panel (c) is disrupted by broadband impulsive RFI at 1.05--1.20~s. This superposition of smearing-induced energy dilution and RFI contamination significantly reduces the local signal-to-interference contrast, limiting the availability of continuous discriminative features and ultimately leading to detection failures.

The missed detections in the $5 \leq \mathrm{S/N} < 10$ range highlight the inherent limitations of deep segmentation models when distinguishing faint signals from residual RFI near the noise floor. Operating directly on raw dynamic spectra places stringent demands on initial local contrast. This underscores a critical need: enhancing the feature representation and structural integrity of weak signals prior to instance segmentation.

\subsection{Validated Parameter Space and Future Improvements}
\label{subsec:parameter_space}

We acknowledge that the current CRAFTS-FRT dataset exhibits class imbalance and limited source diversity. Specifically, it contains only 15 instances from 3 RRATs and 262 instances from 2 repeating FRBs, lacking representation of one-off FRBs and extreme-DM events ($\mathrm{DM} > 1000$~pc~cm$^{-3}$). Given these dataset constraints and the potential for unexpected morphological variations in unseen candidates, we outline the effective parameter space for the current model. Based on our quantitative analysis, \texttt{FRTSearch} demonstrates high reliability for events with $\mathrm{S/N} \geq 10$ and $\mathrm{DM} \lesssim 600$~pc~cm$^{-3}$. When applying this pipeline outside these boundaries---particularly in extreme-DM environments ($\mathrm{DM} > 1000$~pc~cm$^{-3}$) or noise-dominated regimes ($\mathrm{S/N} < 10$)---users should exercise caution and consider cross-validating detections with complementary single-pulse search methods.

Future work will focus on addressing these limitations to further improve the sensitivity and generalization capabilities of \texttt{FRTSearch}. To mitigate the impact of dispersion smearing, we will explore advanced preprocessing strategies, such as more robust RFI mitigation \citep{morello2022iqrm,chen2023rfi} and adaptive contrast enhancement \citep{cao2025basset}, to boost signal-to-background feature separability prior to instance segmentation. Concurrently, we plan to expand the training dataset to encompass greater source diversity, more complex morphologies, and lower-frequency observations (e.g., 400--800~MHz). Ultimately, these enhancements aim to establish \texttt{FRTSearch} as a robust and scalable pipeline for the systematic discovery of astrophysical transients across diverse large-scale radio surveys.

\section{Conclusions}
\label{sec:conclusions}

In this work, we introduced \texttt{FRTSearch}, an end-to-end framework motivated by a fundamental morphological universality: single-pulse emissions from pulsars, RRATs, and FRBs all exhibit consistent dispersive trajectories rigorously governed by the cold plasma dispersion relation (Equation~\ref{eq:dispersion}). By exploiting this intrinsic physical signature, our approach reframes Fast Radio Transient discovery from exhaustive parameter space searches to a direct pattern recognition problem.

\begin{itemize}
    \item We propose a paradigm shift in FRT detection from the computationally expensive ``search-then-identify'' approach to a highly efficient ``detect-and-infer'' model. To facilitate this, we constructed CRAFTS-FRT, a pixel-level annotated dataset derived from the CRAFTS survey \citep{li_2018}. Beyond training our Mask R-CNN model \citep{he2017mask}, this dataset provides a robust foundation for future deep learning applications in radio transient astronomy. Coupled with our novel physics-driven IMPIC algorithm, the framework unifies signal segmentation with the direct inference of DM and ToA, eliminating the need for redundant grid searches.
    
    \item The HRNet-based instance segmentation achieves 77.9\% bbox AP and 57.3\% segm AP on the validation set, comparable to COCO benchmark standards \citep{lin2014}. Furthermore, the IMPIC algorithm inverts segmented trajectory coordinates via the dispersion relation (Equation~\ref{eq:dm}) to deliver DM measurements with a weighted fractional error of 3.8\% across diverse validation sources. This precision is sufficient to generate reliable diagnostic plots for immediate human verification.
    
    \item Benchmarking on the FAST-FREX dataset \citep{guo2024} demonstrates that \texttt{FRTSearch} achieves a recall of 98.0\%, competitive with established exhaustive search methods. Crucially, our approach reduced the generation of false positive candidates by over 92\% compared to Heimdall and TransientX, and by over 99.9\% compared to PRESTO. This dramatic reduction in redundancy, combined with a processing speedup of up to $13.9\times$, significantly alleviates the manual verification workload.
    
    \item The framework exhibited strong generalization capabilities by detecting all 19 tested FRBs from the ASKAP survey \citep{shannon2018} without any retraining. This confirms that the model learns the intrinsic physical signatures of astrophysical sources rather than instrument-specific artifacts.
    
    \item While acknowledging sensitivity limitations for extremely faint signals, \texttt{FRTSearch} offers a scalable, high-precision solution for real-time discovery, establishing itself as a robust tool for processing petabyte-scale data streams from modern radio astronomy facilities like FAST and the SKA.
\end{itemize}

\section*{Acknowledgments}
We thank the referee for helpful comments and constructive suggestions. This work was supported by the National Natural Science Foundation of China (NSFC) Program No. 12588202, the Cultivation Project for FAST Scientific Payoff and Research Achievement of CAMS-CAS (FAST[2019sr04]), the Guizhou Provincial Graduate Research Fund Project (2024YJSKYJJ188), the CAS project No. JZHKYPT-2021-06, the National SKA Program of China (Nos. 2020SKA0120200 and 2022SKA0130104), and the National Key R\&D Program of China (No. 2023YFE0110500). P.W. acknowledges support from the CAS Youth Interdisciplinary Team, the Youth Innovation Promotion Association CAS (id. 2021055), and the Cultivation Project for FAST Scientific Payoff and Research Achievement of CAMS-CAS. D.L. is a New Cornerstone Investigator.

This work made use of data from the Five-hundred-meter Aperture Spherical radio Telescope
(FAST). FAST is a Chinese national mega-science facility, built and operated by the National Astronomical Observatories, Chinese Academy of Sciences.

\section{Data Availability}
The pixel-level annotated CRAFTS-FRT dataset constructed in this work is publicly available at \url{https://doi.org/10.57760/sciencedb.Fastro.00038}. 
The source data for the FRB instances included in this dataset can be accessed via their respective repositories: FRB 20121102 at \citep{li2023} and FRB 20201124A at \citep{zhang2023}. 
Additionally, the FAST-FREX dataset used in this work is available in \citep{guo2024}, and the ASKAP FRB test data are available in \citep{shannon2018}.

\software{ FRTSearch \citep{zhang_2026_18877413}, Astropy \citep{price2022astropy}, PRESTO \citep{ransom2001}, PyTorch \citep{paszke2019pytorch}, MMDetection \citep{chen2019mmdetection}, SciPy \citep{virtanen2020scipy}, Numpy \citep{Harris2020}, Matplotlib \citep{2007Matplotlib},  scikit-image \citep{vanderWalt2014}, pycocotools \citep{lin2014}, your \citep{Aggarwal2020}, tqdm \citep{daCosta-Luis2019}.}

\appendix
\section{Appendix information}

\begin{table}[htbp]
\centering
\small
\setlength{\tabcolsep}{3pt}
\caption{
Detection performance of \texttt{FRTSearch} on 19 ASKAP FRBs. Published parameters from \citep{shannon2018}; inferred DM, ToA, and confidence scores from IMPIC and Mask R-CNN.
}
\label{tab:askap_detection_results}
\begin{tabular}{lccccccc}
\toprule
\textbf{FRB Name} & \textbf{R.A.} & \textbf{Decl.} & \textbf{S/N} & \textbf{Published DM} & \textbf{Inferred DM} & \textbf{Inferred ToA} & \textbf{Confidence} \\
 & \textbf{(hh:mm)} & \textbf{(dd:mm)} & & \textbf{(pc\,cm$^{-3}$)} & \textbf{(pc\,cm$^{-3}$)} & \textbf{(s)} & \\
\midrule
FRB20170416   & 22:13 & $-$55:02 & 8.3  & 523.2  & 494.0  & 3260.7 & 0.58 \\
FRB20170428   & 21:47 & $-$41:51 & 8.2  & 991.7  & 1070.3 & 1893.5 & 0.36 \\
FRB20170707   & 02:59 & $-$57:16 & 10.5 & 235.2  & 239.5  & 570.1  & 0.82 \\
FRB20170712   & 22:36 & $-$60:57 & 13.8 & 312.8  & 310.6  & 2638.1 & 0.97 \\
FRB20170906   & 21:59 & $-$19:57 & 18.6 & 390.3  & 385.1  & 2016.1 & 0.84 \\
FRB20171003   & 12:29 & $-$14:07 & 14.3 & 463.2  & 470.4  & 1164.6 & 0.93 \\
FRB20171004   & 11:57 & $-$11:54 & 11.4 & 304.0  & 272.3  & 624.8  & 0.16 \\
FRB20171019   & 22:17 & $-$08:40 & 24.3 & 460.8  & 467.0  & 1637.9 & 0.94 \\
FRB20171020   & 22:15 & $-$19:40 & 24.4 & 114.1  & 116.9  & 2616.4 & 0.83 \\
FRB20171116   & 03:31 & $-$17:14 & 11.9 & 618.5  & 614.5  & 72.0   & 0.78 \\
FRB20171213   & 03:39 & $-$10:56 & 27.1 & 158.6  & 160.0  & 1938.3 & 0.96 \\
FRB20171216   & 03:28 & $-$57:04 & 8.6  & 203.1  & 186.2  & 1634.5 & 0.37 \\
FRB20180110   & 21:53 & $-$35:27 & 34.7 & 715.7  & 735.6  & 1598.1 & 0.93 \\
FRB20180119   & 03:29 & $-$12:44 & 17.3 & 402.7  & 401.1  & 1679.5 & 0.93 \\
FRB20180128.0 & 13:56 & $-$06:43 & 13.0 & 441.4  & 438.1  & 3348.0 & 0.76 \\
FRB20180128.2 & 22:22 & $-$60:15 & 10.8 & 495.9  & 494.7  & 2719.8 & 0.93 \\
FRB20180130   & 21:52 & $-$38:34 & 10.8 & 343.5  & 347.7  & 572.9  & 0.35 \\
FRB20180131   & 21:49 & $-$40:41 & 14.8 & 657.7  & 658.3  & 16.4   & 0.50 \\
FRB20180212   & 14:21 & $-$03:35 & 18.3 & 167.5  & 167.3  & 1848.6 & 0.98 \\
\bottomrule
\end{tabular}
\end{table}

\bibliography{sample631}{}
\bibliographystyle{aasjournal}



\end{document}